\documentclass[aps,physrev,reprint,superscriptaddress]{revtex4-2}
\usepackage{amssymb}
\usepackage{graphicx}
\usepackage{amsmath,amsfonts}
\usepackage[colorlinks=true,allcolors=blue]{hyperref}
\usepackage{xcolor}  
\usepackage[normalem]{ulem}
\usepackage[T1]{fontenc} 
\usepackage{multirow}

\usepackage{booktabs}
\usepackage{siunitx}
\usepackage{threeparttable}
\usepackage{tabularx}

\usepackage{mathtools}
\usepackage{xcolor}
\usepackage[normalem]{ulem}
\usepackage{url}
\usepackage{subfigure}
\usepackage[colorlinks=true,allcolors=blue]{hyperref}
\usepackage{float}
\usepackage{wasysym}
\usepackage{array}      
\usepackage{longtable}  

\usepackage{amsthm}

\usepackage{algorithm}
\usepackage{algpseudocode}

\algblock{Input}{EndInput}
\algnotext{EndInput}
\algblock{Output}{EndOutput}
\algnotext{EndOutput}
\newcommand{\Desc}[2]{\State \makebox[2em][l]{#1}#2}

\usepackage{comment}

\begin{document}


\title{Efficient generation of networks with minimal average shortest-path distance}

\author{Meritxell Vila-Mi\~nana}
\affiliation{Center for Complex Networks and Systems Research, Luddy School of Informatics, Computing, and Engineering, Indiana University, Bloomington, IN, USA}
\author{Filippo Radicchi}
\affiliation{Center for Complex Networks and Systems Research, Luddy School of Informatics, Computing, and Engineering, Indiana University, Bloomington, IN, USA}


\begin{abstract}
Designing networks that minimize distances and satisfy structural constraints is a fundamental task across transportation, communication, and biological systems. Here, we consider the problem of 
finding, for a given degree sequence, the network structure displaying the smallest possible average shortest-path length. 
While exact solutions are available in linear time for trees, such an optimization problem becomes computationally infeasible as soon as loops are allowed in the networks. 
We propose a fast algorithm to construct approximate solutions to such a degree-constrained distance-minimization problem. Accordingly, edges are first created between high-degree nodes; then, additional connections are placed following the rules of the standard configuration model. In spite of its simplicity, the algorithm displays outstanding performance as demonstrated in our systematic experiments on both synthetic and real degree sequences.
Our method is particularly effective on synthetic degree sequences displaying medium levels of heterogeneity. When applied to degree sequences of real networks, the proposed algorithm is able to reduce the all-pair shortest path of real structures by 20\%, on average. 
We perform a validation on
small-sized networks, where we compare the shortest-path distance of the networks generated with our algorithm against those obtained via simulated annealing optimization.
Although simulated annealing yields slightly better structures, our proposed algorithm provides nearly identical  solutions at a substantially lower computational cost, 
making it a solid method in applications concerning large-scale systems.
\end{abstract}

\maketitle
\let\oldaddcontentsline\addcontentsline
\renewcommand{\addcontentsline}[3]{}

\section{Introduction}

Consider an already built airport infrastructure. Each airport can support a 
certain
number of flights. 
How should we arrange flights between airports to allow people taking the least amount of flights between their demanded origin-destination pairs?
Similarly in a communication network where each node has a prescribed number of available connections, how do we establish which pairs of nodes should be connected so that information could travel through the smallest amount of intermediate nodes?
More in general, is there a way of determining, for a given degree sequence, the network structure displaying
the smallest possible average shortest-path length?

As apparent from Refs.~\cite{goubko2015minimizing, lin2013extremal, furtula2013more},
such a constrained optimization problem
is fully understood in the case of trees, i.e., networks without loops.
These papers focus on the minimization of the Wiener index~\cite{wiener1947structural}, which is the sum of all-pair shortest path distances, thus equal to the average path distance except for a constant multiplicative factor. The Wiener index was originally developed
to characterize chemical molecules based on their structure~\cite{wiener1947structural, rouvray1976dependence, gutmana1995wiener, stiel1962normal}, and its mathematical properties 
have been widely studied~\cite{knor2015mathematical, dobrynin2001wiener}.
On trees, 
the degree-constrained network optimization problem is exactly solvable with an algorithm having linear time complexity \cite{goubko2015minimizing}. Also, it has been shown that a balanced tree (also known as a Volkmann tree) is the one that minimizes the average shortest-path length across the set of all trees with a fixed maximum node degree \cite{fischermann2002wiener}. For trees with fixed number of nodes, minimizers 
of the average shortest-path length have been fully characterized in Refs.~\cite{lin2013extremal, furtula2013more}. 
In particular, it has been proved that among all trees with 
even number of 
vertices, where all vertices have odd degree, the star 
uniquely minimizes the average shortest path.

However, the minimization of the average shortest path for degree sequences that lead to the presence of loops has been overlooked. 
Only a few results are known, but with little practical relevance. First, it is known that
exact solutions can be obtained in polynomial time only for degree sequences admitting at most one loop, whereas, in general settings, the problem is computationally unfeasible~\cite{goubko2015minimizing}. Also, the generalization from trees to loopy networks is far from being trivial, as even the addition of just a loop may dramatically change the structure of the optimal network with respect to the one obtained in absence of such a loop~\cite{burger2024minimizing}.

\begin{figure}[h]
    \centering
\includegraphics[width=0.94\columnwidth]{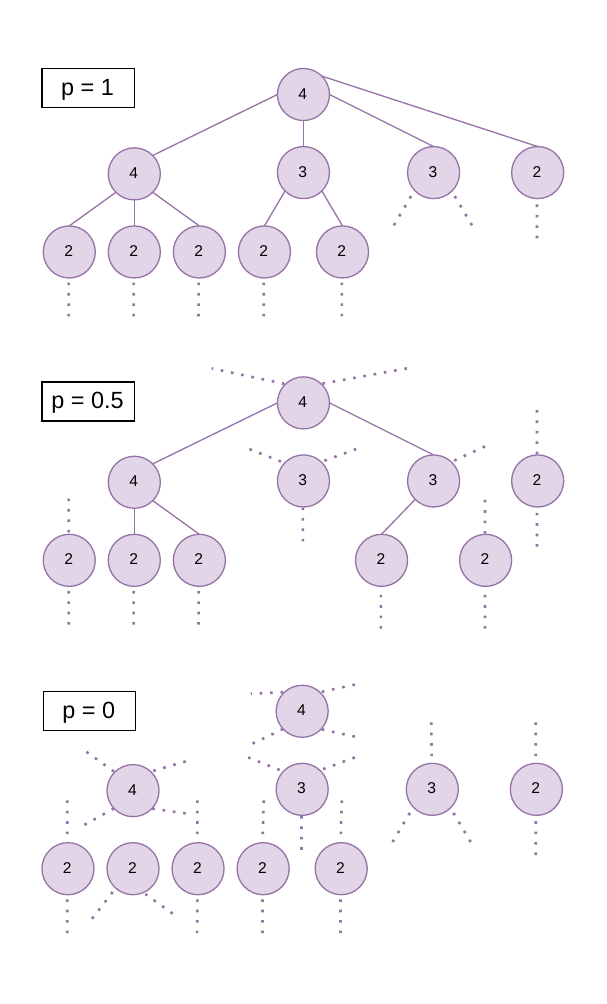}
\caption{Illustration of the degree-biased configuration (DBCM) model. The algorithm takes as input a degree sequence to construct a network in two phases: (i) a greedy-tree (GT) construction where high-degree nodes are preferentially attached to form a loop-less structure (ii) followed by a configuration-model (CM) phase acting on the remaining degree sequence. The parameter $0 \leq p \leq 1$ tunes the importance of one phase over the other. In the illustration, segments denote edge constructed during the GT phase of the algorithm, whereas dashed lines represent stubs to be paired using the CM. For $p=1$, the GT construction is deterministic, so that all high-degree nodes exhaust their available connections in the GT phase. For $p=0$, no edges are formed in the GT and all degrees remain available for the CM phase. For $p=0.5$, part of the hierarchical structure is constructed, while some intermediate nodes have available degrees for the CM phase. }
\label{fig: p_greedy_graph_construction}
\end{figure} 

In this paper, we 
study the optimization problem for networks with realistic degree sequences, thus potentially admitting many loops.
To this end, we develop 
an algorithm,
which we call 
degree-biased configuration model (DBCM)
aimed at approximating solutions of the degree-constrained minimization problem of the average shortest-path length.
DBCM is specifically conceived for degree sequences of large-scale, sparse, tree-like networks, combining the method by Goubko {\it et al.}, designed for trees, and the configuration model (CM) by Molloy and Reed, traditionally used to generate random graphs with given degree sequences~\cite{goubko2015minimizing, molloy1995critical}. In the first phase of the algorithm, a loop-less structure, where edges are preferentially established between pairs of high-degree nodes, is created (see Figure~\ref{fig: p_greedy_graph_construction}); in the second phase, loops are introduced in the network as the remaining connections are established according to the rules of the CM. The relative importance of one phase over the other is tuned based on an input parameter. We systematically test the performance of DBCM on both synthetic and real degree sequences against different types of baselines. On relatively small networks, we compare DBCM against simulated annealing (SA) optimization, showing that DBCM solutions are almost identical to the SA ones. On real degree sequences, we show that DBCM reduces the average shortest-path length of the corresponding real networks by about 20\%. Finally, in all networks we compare DBCM against the standard CM. Our main finding that DBCM consistently outperforms the standard CM indicates that the biased creation of connections between high-degree nodes is key to obtain network structures characterized by small shortest-path distances.

\section{Problem formulation}\label{sec:problem-formulation}


Let $G$ be an undirected multi-graph, thus admitting self-loops and parallel edges, composed of $N$ nodes. We assume that the topology of $G$ is fully specified by its adjacency matrix whose generic element $A_{n,m} \in \mathbb{N}$ directly quantifies the number of parallel edges existing between nodes $n$ and $m$. Since the graph is undirected, we have $A_{n,m} = A_{m,n}$ for all $n, m = 1, \ldots, N$. 
The degree of node $n$ is defined as $k_n = \sum_{m=1}^N A_{n,m}$.
The average shortest-path length $\langle \ell (G) \rangle$ of 
the
graph $G$ is defined as
\begin{equation}
\langle \ell (G) \rangle = \frac{2}{N(N-1)} \sum_{n = 1}^{N-1}  \sum_{m = n+1}^N \, \ell (n,m) \; ,
\label{eq:wiener}
\end{equation}
where
$\ell(n,m)$ denotes the length of the shortest path between vertices $n$ and $m$, 
i.e., the minimum number of edges that must be traversed to go from $n$ to $m$ in the graph $G$.
Since the graph is symmetric, $\ell(n,m) = \ell(m,n)$; also, we  have $\ell(n,n) = 0$. Two additional remarks are in order. First, we assume that the length of the shortest path between two nodes is not affected by the presence of parallel edges along the path. Thus, even if we allow for the presence of self-loops and parallel edges, optimal structures are preferentially simple graphs. Second, we use the convention that, if nodes $n$ and $m$ are part of two different connected components of the graph $G$, then $\ell(n,m) = \infty$. 
This fact implies that,
when seeking graph structures with minimal average shortest-path length, the search is {\it de facto} restricted to networks formed  by a single connected component only. 

We focus our attention only on graphs $G$ with specified degree sequence $\mathbf{k} = (k_1, \dots, k_N) \in \mathbb{N}_+^n$, with $\sum_{n=1}^N k_n$ even.
We indicate the set of such graph as $\mathcal{G}(\mathbf{k})$. Within the set, we identify as the optimal graph, namely $G^*(\mathbf{k})$, as the one that corresponds to the smallest value of the average shortest-path distance, i.e., 
\begin{equation}
G^*(\mathbf{k}) = \arg \min_{G \in \mathcal{G}(\mathbf{k})} \langle \ell(G) \rangle \; .
\label{eq:wiener-min}
\end{equation}

\section{Methods}
\subsection{Exact solution for trees}\label{sec: Solution for trees}

It has been shown independently by Wang \cite{wang2008extremal} and Zhang {\it et al.}~\cite{zhang2008wiener} that the so-called greedy tree (GT) is the graph structure that minimizes the average shortest-path length among all trees with a fixed degree sequence. Being a tree,
the input is a degree sequence $\mathbf{k} = (k_1, \ldots, k_N)$ such that
\begin{equation}\label{eq: generating tuple}
    \sum_{n = 1}^N k_n = 2(N - 1).
\end{equation}
The GT is built with a top-down strategy, starting with the vertex with maximum degree in the graph, which becomes the root of the tree. Then, at each step, the remaining vertices are selected in descending order of degree, and are attached greedily to the tree, so that each selected vertex is connected to an existing vertex, that still has available capacity. This way, 
high-degree
vertices are placed 
closer to the root, 
and the tree spreads out its branches sooner, which helps to keep it balanced. Similarly, the vertices with fewer connections are added later and usually end up at the outer parts of the tree.


This approach was generalized by Goubko~\cite{goubko2015minimizing}, who provided an exact solution to the minimization problem of weighted trees, under the assumption of degree-monotone node weights, through a construction that extends the Huffman's algorithm~\cite{huffman1952method}. When all weights are identical, this construction corresponds to the GT's one. 

\subsection{Exact solution for unicyclic networks}\label{sec: Solution for trees}

Exact solutions  to the problem of Eq.~(\ref{eq:wiener-min}) can be found in polynomial time also for degree sequences such that $\sum_{n=1}^N k_n = 2 N$, thus leading to networks that admit one cycle~\cite{burger2024minimizing}. 
An important caveat is that the optimization problem of Eq.~(\ref{eq:wiener-min}) is constrained not just by the input degree sequence, but also by the girth (i.e., the length of the shortest cycle) of the resulting graph.
In particular, Burger and Rakotonarivo have proved that any minimizer 
must belong to one of these three families: (i) the \textit{greedy unicyclic graph}, where the largest degrees are placed as centrally as possible on the cycle and the remaining vertices are attached in decreasing order; (ii) the \textit{cycle-centered graph}, where the largest degrees are assigned to the cycle vertices and the remaining vertices are arranged to satisfy the degree sequence; (iii) the \textit{out-greedy unicyclic graph}, obtained by first constructing a GT and then adding edges to create the prescribed cycle away from the centroid, namely the set of vertices minimizing the total distance to all other vertices.
Consequently, solving Eq.~(\ref{eq:wiener-min}) reduces to evaluating the average shortest-path distance of these three candidate constructions and selecting the one with the smallest value.
In other words, the presence of even a single loop makes the optimization problem of Eq.~(\ref{eq:wiener-min}) more challenging than in loop-free structures. This observation highlights the inherently nontrivial nature of the problem for networks with arbitrary degree distributions, which we investigate next.

\subsection{Approximating solutions in arbitrary graphs via the configuration model}\label{subsec:conf_model}

A natural choice to construct networks starting from their degree sequence is the configuration model (CM)~\cite{molloy1995critical}.
CM is the standard generative model in network science to create graphs with random structure that are solely constrained by their degree sequence~\cite{newman2010networks}.
The procedure of construction of the CM is very simple. Given a degree sequence $\mathbf{k} = (k_1,\dots,k_N)$, the CM constructs a random multigraph as follows. For each vertex $n$, one creates $k_n$ half-edges, or stubs, which are then paired uniformly at random to form edges. The resulting graph satisfies the degree sequence by construction, and it may produce self-loops and multiple edges.
Also, the resulting graph can be composed of more than one component.

If connected, CM-generated graphs typically display small average shortest-path distances, thus they provide strong baselines to compare against~\cite{newman2010networks}. For example, if degrees are random variates obeying a power-law degree distribution $P(k) \sim k^{-\lambda}$ with exponent $\lambda > 2$, the average shortest-path distance displays a scaling that depends on the network size $N$ in a specific manner: if  $\lambda > 3$, the degree distribution has finite variance and the average shortest-path distance scales as $\mathcal{O}(\log N)$, as in small-world networks~\cite{watts1998collective}; if instead $2 < \lambda \leq 3$, the variance diverges and distances shrinks to $\mathcal{O}(\log\log N)$, giving rise to the so-called ultra-small world phenomenon \cite{chung2002average,cohen2003scale}.
This distinction is particularly relevant in our work. For power-law sequences with $2 < \lambda \leq 3$, the configuration model already produces extremely short distances due to the presence of hubs. Since distances are already of double-logarithmic order, it may be more difficult for any other graph structure to produce extra improvement. In contrast, when $\lambda > 3$, distances scale logarithmically, leaving more room for structural optimization.

\subsection{Approximating solutions in arbitrary graphs via the degree-biased configuration model}

Learning from the well-established construction procedures that we just described,  we introduce the degree-biased configuration model (DBCM), a model for the construction of networks that combines features from both the GT and CM generative algorithms.

To create an instance of the DBCM, 
we begin by constructing a stochastic version of GT. Let $\mathbf{k} = (k_1, \dots, k_N)$ be the degree sequence sorted in decreasing order, so that $k_1 \geq k_2 \geq \dots \geq k_N$. The vertex with largest degree is selected as the root. Next, the remaining vertices are stored in a max-heap ordered by available degree, and we keep track on a queue of the vertices that were already inserted into the tree and still have available degree. Iteratively, we extract a vertex $u$ from the queue and attempt to attach to it the node $v$ from the heap with highest available degree. The edge $(u,v)$ is added with probability $p \in [0,1]$. If the edge is accepted, the residual degrees of $u$ and $v$ are decreased, and if $v$ still has available degree, it is appended to the queue. This process continues until no further attachments are possible. 
After that, edges connecting nodes with remaining available degree are created using the classical CM, as described in Section~\ref{subsec:conf_model}. Algorithm~\ref{alg:stochastic_greedy_conf} in Appendix~\ref{sec:Implemented_algorithms} summarizes our implementation of the DBCM.

Note that when $p=1$, the construction of the tree structure is equivalent to GT. Loops are then created, but mostly involving low-degree nodes. For $p=0$, no edges are created in the greedy phase and the entire graph is generated according to the CM's rules.  For intermediate $p$ values, only part of the hierarchical GT structure is actually created during the GT construction; this choice allows cycles to close, in the CM phase, between nodes away from the leaves of the GT, an operation that resembles the cycle-centered or out-greedy constructions of Ref.~\cite{burger2024minimizing}. In short, through the tunable parameter $p$, one can continuously go from the pure CM ($p=0$) to the deterministic GT backbone ($p=1$). Figure~\ref{fig: p_greedy_graph_construction} shows a schematic representation of the stochastic tree obtained after the first phase of the algorithm, before completing the graph with the CM phase. The figure illustrates the deterministic GT construction for $p=1$, one possible realization of the stochastic tree for $p=0.5$, and the initial disconnected state for $p=0$. Dashed lines indicate remaining stubs to be paired in the CM phase.

\subsection{Approximating solutions in arbitrary graphs via simulated-annealing optimization}
\label{sec: SA solution}

Approximate solutions to the problem of Eq.~(\ref{eq:wiener-min}) can be found via
simulated annealing (SA) optimization, a traditional optimization technique for problems with rugged energy landscapes \cite{kirkpatrick1983optimization}. 
SA is known to generate solutions that are quasi optimal by, however, incurring a substantial computational cost. This restricts the application of SA to relatively small systems only.

For our problem of Eq.~(\ref{eq:wiener-min}), each graph $G \in \mathcal{G}(\mathbf{k})$ is assigned with an energy equal to its average shortest-path distance of Eq.~(\ref{eq:wiener}).


An important component of simulated annealing is the definition of a neighborhood relation that allows us to explore the feasible graphs in the state space. In our problem, candidate neighbor graphs must preserve the degree sequence $\mathbf{k}$ and the connectedness of the graph. Both these features are imposed in the initial graph generated using the rules of the configuration model~\cite{molloy1995critical}.

At each stage of the SA algorithm,
we obtain a neighbor of the current graph $G$ by rewiring two edges.
Specifically, we obtain a neighbor graph $G'$ of $G$ by selecting two distinct edges $(u,v)$ and $(w,z)$ from $G$, removing them, and reconnecting the four endpoints with either $(u,z),(w,v)$ or $(u,w),(z,v)$. 
This procedure ensures that the degree sequence is preserved.


Given a current graph $G$ and a candidate neighbor $G'$, the change in the objective function is
$ \langle \ell(G') \rangle - \langle \ell(G) \rangle$
Then, the candidate graph is accepted with probability given by
\begin{equation}
\label{eq:prob_accept}
P(G \to G') =
\min\left\{1,\; e^{-\frac{\langle \ell(G') \rangle - \langle \ell(G) \rangle}{T}} \right\},
\end{equation}
where $T > 0$ represents the current temperature.
Hence, we always accept improvements, i.e., $\langle \ell(G') \rangle \leq \langle \ell(G) \rangle$, while worse solutions may be accepted with a probability that decreases both with 
the energy differential and the temperature parameter.
This mechanism allows the algorithm to escape local minima during the early stages of the search. Note that, since the distance between two nodes belonging to disconnected components is infinite, the probability of accepting of a graph $G'$ that is not composed of just one connected component is equal to zero. This ensures that the connectedness of the graph is preserved during SA exploration.

In our SA optimization protocol, 
we start from the temperature $T_0 = 0.1$. At each stage $s$, we consider $60$ connected candidate graphs, eventually accepting changes in the structure according to Eq.~(\ref{eq:prob_accept}). We then move to the next stage by decreasing the temperature as $T_{s+1} = \alpha \, T_s$,
with a cooling rate $\alpha = 0.9$, $0 < \alpha < 1$. The optimization algorithm ends when we reach the temperature 
 $T_{\min} = 0.001$.

 \subsection{Approximating the average shortest-path length of a network}
 \label{sec:wiener_approx}

Computing the exact value of the average shortest-path distance of Eq.~(\ref{eq:wiener}) requires estimating the pairwise distance between all pairs of nodes. Its time complexity scales rapidly; it is $\mathcal{O}(NM)$ for sparse graphs, where $N$ is the number of nodes, and $M$ is the number of edges.

To reduce the computational cost, we uniformly sample $Z$ root nodes and, for each root 
we compute single-source shortest-path distances using breath-first search (BFS). Each BFS costs $\mathcal{O}(N+M)$, so the time required for the measurement scales as $\mathcal{O}(Z(N+M))$. We use the mean of the sampled distances as an estimator of the average shortest-path length.


We validate the accuracy of the sampling estimator by comparing it against the exact average shortest-path length on small networks, where exact computation is feasible. We observe a clear convergence as $Z$ increases. In particular, for $Z = 20$, the relative error is below 2\% for all network sizes, which indicates that the sampling estimator provides a reliable approximation of the average shortest-path distance (see Figure~\ref{fig:sampling_validation} in Appendix~\ref{sec:appendix_avg_sht_path_sampling_k_roots}). 

\section{Results}

\subsection{Synthetic power-law degree sequences}


We test the above algorithms
on degree sequences
randomly generated
from the power-law distribution 
    $P(k) \sim k^{-\lambda}$ if $k \in \left[x_{\min}, k_{\max} \right]$, and $P(k) = 0$ otherwise. We set $k_{\max} = \min\{\sqrt{N}, N^{1/(\lambda-1)}\}$, as generally imposed for the generation of uncorrelated CM networks~\cite{catanzaro2005generation}.
    We set $k_{\min}=3$ in most of our experiments, except for a few where we set $k_{\min} =5$.

\begin{figure*}[!htb]
    \centering
\includegraphics[width=\textwidth]{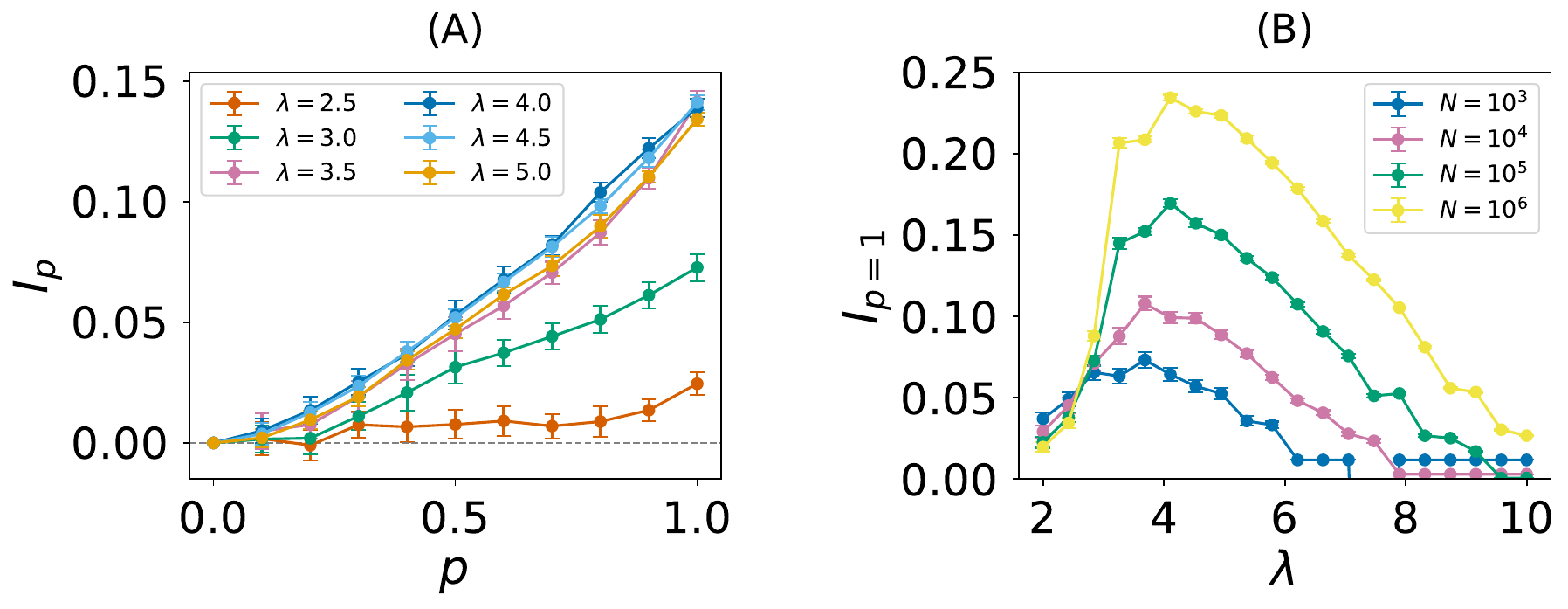}
\caption{(A) Relative improvement $I_p$ of the DBCM construction with $p$ over the $p=0$ case as a function of the parameter $p$, for networks with $N = 10^5$ nodes, different $\lambda$ values and minimum degree $k_{\min}=3$. Results are averaged over 20 degree sequences, with 10 graph realizations per sequence and per value of $p$. We display with symbols the average value of $I_p$; error bars denote the standard deviation of $I_p$. (B) Relative improvement $I_{p=1}$ of the DBCM construction with $p=1$ over $p=0$ as a function of the power-law exponent $\lambda$, for networks of varying sizes $N$ and minimum degree $k_{\min}=3$. Results are averaged over 50 independent degree sequences for each $\lambda$ and 50 independent graph realizations of parameter $p$ for each degree sequence. 
}
\label{fig:lambda_rel_improvement}
\end{figure*}

Results of our experiments are summarized in Figure~\ref{fig:lambda_rel_improvement}.
As a main evaluation metric, we use the relative improvement of the objective function of the DBCM with respect to the CM performance. This is defined as
\begin{equation}\label{eq: rel_imp_av}
I_p = \frac{\langle \ell \rangle_{p=0} - \langle \ell \rangle_{p}}{\langle \ell \rangle_{p=0}} \; ,
\end{equation}
where $\langle \ell \rangle_p$ is the actual value of the average shortest-path length obtained using the DBCM algorithm for given value of the parameter $p$. Here, the average shortest-path length is approximated using $Z = 50$ root nodes, as described in Sec.~\ref{sec:wiener_approx}.


Figure~\ref{fig:lambda_rel_improvement}A shows $I_p$ as a function of $p$ for different values of the power-law exponent $\lambda$.
Here, the system size is kept constant to $N = 10^5$. We observe that in all cases the improvement increases consistently with $p$. In particular, $p=1$ always yields the lowest value of the average shortest-path length. This indicates that favoring the initial tree construction, which attaches high-degree nodes first, systematically reduces the average shortest-path length. 

Moreover, the improvement is substantially more pronounced for $\lambda=4.0$ than for $\lambda=2.5$. 
In Figure~\ref{fig:lambda_rel_improvement}B, we plot $I_{p=1}$ as a function of the degree exponent $\lambda$. We observe a 
non-monotonic behavior, with a clear peak reached at around $\lambda = 4.0$; the peak becomes more pronounced as the system size increases, reaching even $I_{p=1} \simeq 0.2$ for $N = 10^6$. Results reported in the Appendix also tell us that the dispersion of the distribution of the shortest-path length is proportional to its average value, see Figure~\ref{fig:power_law_histograms}. Also,   results obtained for $k_{\min} = 5$, and for the relative improvement of the standard deviation, confirm the same type of qualitative behavior, see Figures~\ref{fig:lambda_rel_improvement_xm3_std} and \ref{fig:lambda_rel_improvement_xm5}.

The finding can be understood by examining how the degree distribution influences the DBCM construction procedure. When $\lambda \simeq 2$, the degree distribution is extremely heterogeneous. A small number of hubs carry a large fraction of the total degree, while most nodes have very small degree. 
In this regime, there are a few hubs that can connect to a very large number of nodes. Pairs of such hubs are likely to be connected by an edge in the CM phase, making the GT phase almost irrelevant for the resulting network structure.  

For intermediate $\lambda$ values, say $3 \leq \lambda \leq 5$, degree heterogeneity is still present, but hubs do not longer dominate the entire degree distribution. In this case, there are multiple nodes with moderately high degree. The CM connects stubs randomly, so edges between high-degree nodes are not necessarily prioritized and occur by chance. Since the $p=1$ construction starts with a hierarchical tree structure, 
it favors connections that shorten distances between high and medium degree nodes early in the process. 
Such a backbone serves to greatly reduces the average distance between nodes in the network compared to what achievable via the CM phase only.

However, as $\lambda$ increases further, the degree distribution becomes increasingly homogeneous. Since no neat hierarchy is present in the degree sequence, the CM and the DBCM constructions lead to very similar results. 

To further characterize the structural organization induced by the $p=1$ construction, we analyze the evolution of degree correlations during the network-assembly process. Figures~\ref{fig:assortativity_plots_node_degree} and \ref{fig:assortativity_plots_num_edges} in Appendix~\ref{subsec_appendix: assortativity} report the average degree of neighbors for the $p=0$ and $p=1$ constructions across synthetic power-law networks with different degree exponents $\lambda$. These results provide additional insight into how the greedy attachment mechanism progressively generates highly disassortative structures, especially for heterogeneous degree distributions with small or medium $\lambda$ values. The supplementary analysis also illustrates how the assortative structure evolves dynamically as edges are added during the graph construction.


We conclude this section by reporting on an empirical analysis of the time complexity of the DBCM algorithm. 
The time $T(N)$ required to generate a graph using DBCM and to estimate the associated average shortest-path length depends on the system size $N$ according to the relation $T(N) \sim N^{1.07}$  (see Figure~\ref{fig:time_complexity} in Appendix~\ref{sec: appendix_time_complexity}). In particular, the generation of the graph requires a time scaling as $T_{\text{gen}}(N) \sim N^{1.05}$, whereas the estimate of the average shortest-path length requires a time scaling as  $T_{\text{est}}(N) \sim N^{1.15}$.
 This denotes a scaling close to linear over the tested range of network sizes, which is consistent with the fact that both network generation and estimation of the average shortest-path length require linear time in sparse networks.
 The procedure 
 can be thus applied to large-scale 
 degree sequences.

\subsection{Optimization of real networks}


We evaluate our proposed DBCM model construction on $109$ real-world networks from different domains, with sizes ranging from $N = 30$ to $N = 5 \times 10^6$. 
In Table~\ref{tab:real_networks} of Appendix~\ref{sec: appendix_table_real_networks}, we provide a detailed description of the real-world networks used in this study, including their number of nodes and edges, and their category (biological, social, communication, citation, information, technological, infrastructure or transportation).
For each dataset, we first preprocess the graph by converting it to an undirected network, if necessary, and we extract its largest connected component. We estimate the average shortest-path length of the graph $W_{\text{real}}$ using the largest connected component only; the degree sequence of such a largest connected component is then used as an input for all algorithms.

For each network  dataset and each value of the parameter $p$, we generate $20$ independent realizations of the DBCM. 
The average shortest-path length of each network is approximated using the method described in Sec.~\ref{sec:wiener_approx} with  $Z = 20$.
While the DBCM construction creates connected graphs for the vast majority of the synthetic degree sequences we considered, the same is not true when applied to degree sequences obtained from real networks. This is especially true for degree sequences with minimal degree $k_{\min} = 1$. Thus, to have a fair comparison across methods, we restrict to networks for which at least one connected realization is obtained for every $p$ value. In Appendix~\ref{appendix: connectivity_prob}, we analyze the probability that the DBCM construction yields a connected graph as a function of $p$. 
For each network, we identify the value of the parameter $p$ that achieves the best performance in terms average value of the average shortest-path length, see Table~\ref{tab:win_counts_real_networks}.  We also include in the comparison the real network itself.
Results are computed over the subset of $39$ networks for which all values of $p$ yield at least one connected realization.

\begin{table}[!htb]
\centering
\vspace{0.3em}
\begin{tabular}{
l@{\hspace{1.5em}}
S[table-format=2.0]
S[table-format=2.1]
S[table-format=2.0]
S[table-format=2.1]
}
\toprule
\textbf{Method} & {\textbf{Count}} & {\textbf{\%}} 
\\
\midrule
\textit{Real}    & 3  & 7.7  \\ 
$p=0.0$   & 0  & 0.0  \\ 
$p=0.1$ & 0  & 0.0  \\ 
$p=0.2$ & 1  & 2.6  \\ 
$p=0.3$ & 1  & 2.6  \\ 
$p=0.4$ & 2  & 5.1  \\ 
$p=0.5$ & 1  & 2.6  \\ 
$p=0.6$ & 1  & 2.6  \\ 
$p=0.7$ & 3  & 7.7  \\ 
$p=0.8$ & 3  & 7.7  \\ 
$p=0.9$ & 1  & 2.6  \\ 
$p=1.0$   & 23 & 59.0 
        \\ 
\bottomrule
\end{tabular}
\caption{Number and percentage of real networks for which each method achieves the minimum value of the average shortest-path length. Results are computed over networks for which all values of the model parameter $p$ yield at least one connected network.}
\label{tab:win_counts_real_networks}
\end{table}

From Table~\ref{tab:win_counts_real_networks}, we observe a clear dominance of the DBCM model for $p=1$, which achieves the lowest average shortest-path length  for 59\% of the networks considered. 
It is interesting to notice that the real networks themselves are optimal in only 7.7\% of the cases. This suggests that real-world networks are typically not arranged to minimize global path length under the constraint of a fixed degree sequence. Moreover, $p=0$, corresponding to the CM case, rarely achieves optimal performance. 

\begin{figure*}[t]
    \centering
\includegraphics[width=\textwidth]{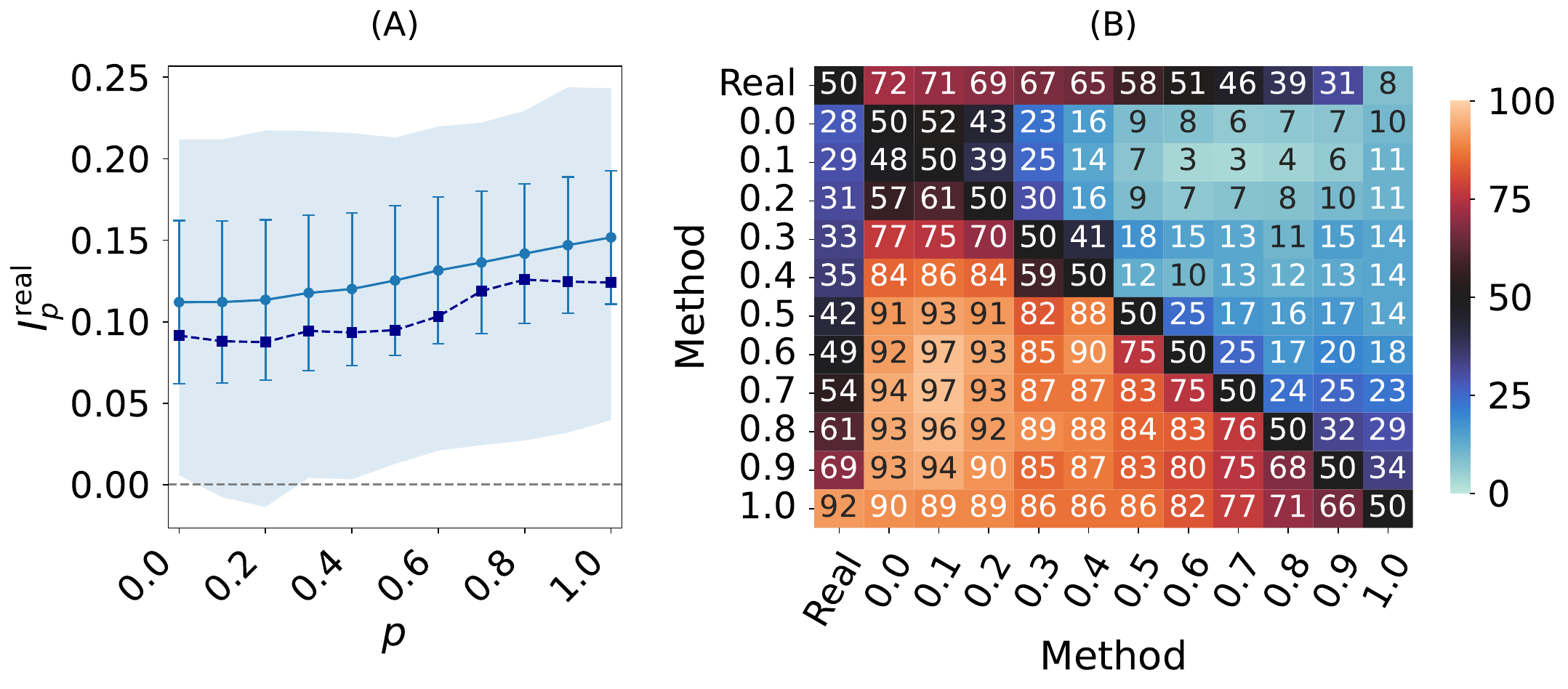}
\caption{(A) Relative improvement of the average shortest-path length obtained using the DBCM with parameter $p$ with respect to the corresponding value observed in the real networks as a function of $p$. Circles denote the mean value of $I^{\text{real}}_p$ and 95\% confidence intervals (CI) are denoted by the corresponding error bars, the dashed line indicates the median of $I^{\text{real}}_p$, while the shaded region represents the interquartile range (IQR). Results are computed over networks for which all $p$ yield at least one connected realization. (B) Pairwise win percentage matrix for different methods. Here, one method ``wins'' over the other if, for a given degree sequences, it generates a network with average shortest-path length smaller than the other method. Ties are counted as half a win for each method. \textit{Real} denotes the original network.
Results shown are for all 109 networks, where disconnected realizations are treated as automatic losses.}
\label{fig:pair_wins_real_networks}
\end{figure*}

Similarly, we analyze the relative improvement 
\begin{equation}
    I^{\text{real}}_p = \frac{\langle \ell \rangle_{\text{real}} - \langle \ell \rangle_{p}} { \langle \ell \rangle_{\text{real}} }
    \label{eq:improv_real}
\end{equation} of the average shortest-path length obtained using the DBCM with parameter $p$ with respect to the metric value observed in the real networks. 
As Figure \ref{fig:pair_wins_real_networks}A,
 $I^{\text{real}}_p$ increases consistently with $p$, indicating that including the initial tree structure  systematically reduces the average distance of the resulting network. Even for small values of $p$, the improvement is consistently positive across most networks with respect to the original real network. In addition, the relatively narrow confidence intervals suggest that the improvement is robust. 
Moreover, we find that the probability of generating a connected network increases with $p$ (see Figure~\ref{fig:connectivity_probability} in Appendix~\ref{appendix: connectivity_prob}).

In Figure~\ref{fig:pair_wins_real_networks}B, 
we perform a pairwise comparison between different $p$ values in the DBCM. We specifically measure the fraction of networks for which one parameter value generates networks with smaller average shortest-path length than those obtained by setting the other parameter value.
We include all 109 real networks; disconnected realizations are treated as automatic losses. This choice penalizes parameter values that frequently fail to produce connected graphs, and therefore captures both performance and robustness. We observe that large values of $p$ tend to dominate small values, 
with $p=1$ outperforming all other methods in the majority of networks, and being particularly strong against the real network itself, in over $90\%$ of the cases. Overall, 
these results indicate that 
increasing the greedy bias $p$ consistently improves performance and robustness.

\begin{figure}[!htb]
    \centering
\includegraphics[width=0.5\textwidth]{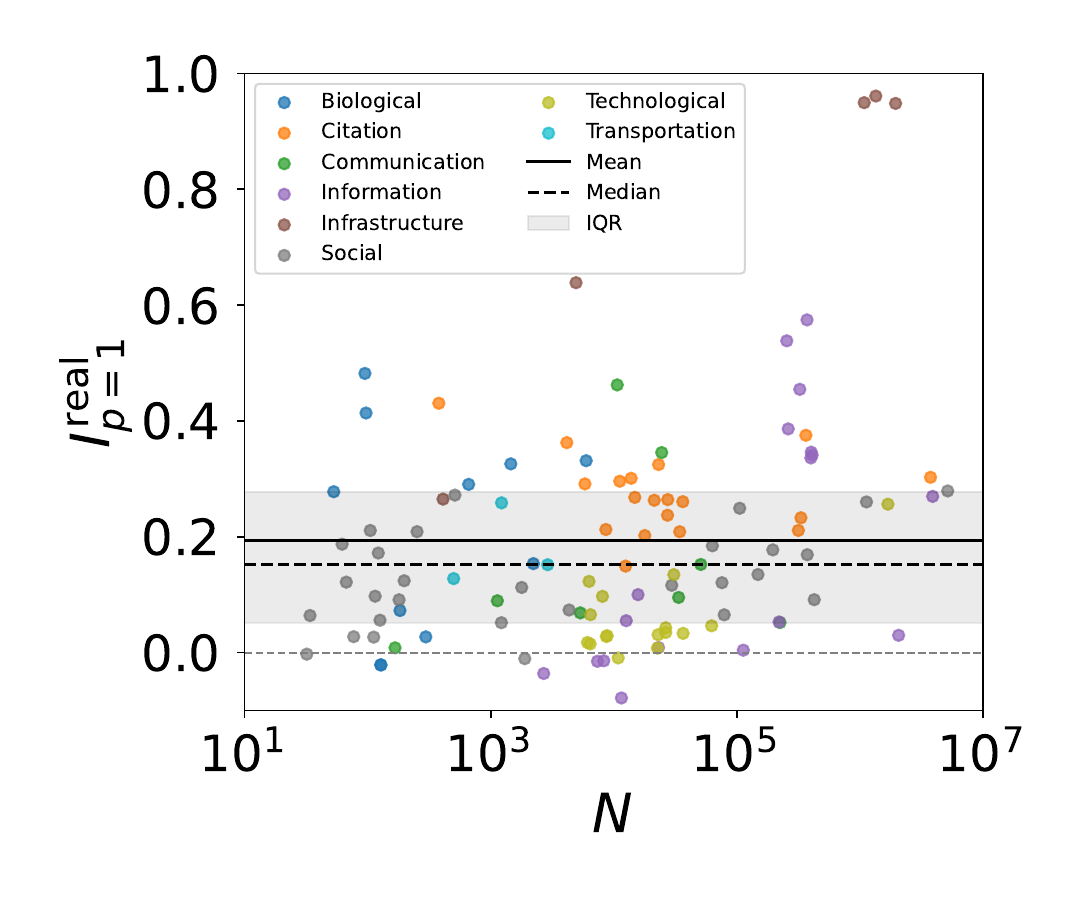}
\caption{Relative decrease of the average shortest-path length achieved by the DBCM with $p=1$ with respect to the real network as a function of the network size. Each point corresponds to a network and is colored according to its category shown in Table~\ref{tab:real_networks}. 
Error bars have size comparable to symbols, so they are not displayed in the figure.
The solid and dashed horizontal lines denote the mean and median improvement across networks, respectively, while the shaded region indicates the interquartile range (IQR).}
\label{fig:rel_improv_vs_size_p1_categories}
\end{figure}

Since $p=1$ is the parameter value yielding the overall best performance by DBCM, and it also guarantees connectivity, we further analyze its performance across all 109 real-world networks. In Figure~\ref{fig:rel_improv_vs_size_p1_categories}, we plot $I^{\text{real}}_{p=1}$ {\it vs.} $N$ for all $109$ networks in our corpus. 
Each point corresponds to a network, colored by its category. 
The improvement is consistently positive for nearly all networks. This indicates that the greedy construction systematically reduces path lengths under the degree sequence constraint. Moreover, the magnitude of the improvement, of around 20\%, is stable across several orders of magnitude in network size. This suggests that the effectiveness of the method does not strongly depend on system size. Overall, this confirms that the $p=1$ construction provides a robust and scalable baseline for minimizing path lengths in real networks.
A special case is given by three large road networks, where $I^{\text{real}}_{p=1}  \simeq 100\%$. However, this is the consequence of the fact that DBCM is not accounting for the spatial distribution of nodes in the real network, thus allowing for unrealistic shortcuts between nodes that are far away in physical space.

To validate the performance of DBCM, 
we extend our analysis by considering a subset of $24$ small real-world networks with up to $N = 500$ nodes, for which SA optimization is computationally feasible. We compare DBCM solutions directly against SA solutions.
For this subset of small networks,
the probability that DBCM leads to connected graphs is significantly higher than in the full dataset, exceeding 90\% even for $p=0$ and increasing to 100\% for $p=1$, see Figures~\ref{fig:connectivity_probability} and~\ref{fig:sa_connectivity} in Appendix~\ref{appendix: small real networks}. Also on this subset, we find that the best performance for DBCM is achieved by setting $p=1$, see Figures~\ref{fig:sa_rel_improv} and~\ref{fig:sa_pairwise} in Appendix~\ref{appendix: small real networks}.
SA solutions are generally better than for DBCM, but the performance gap betweent he two methods is relatively small.

\begin{figure}[!htb]
    \centering
\includegraphics[width=0.5\textwidth]{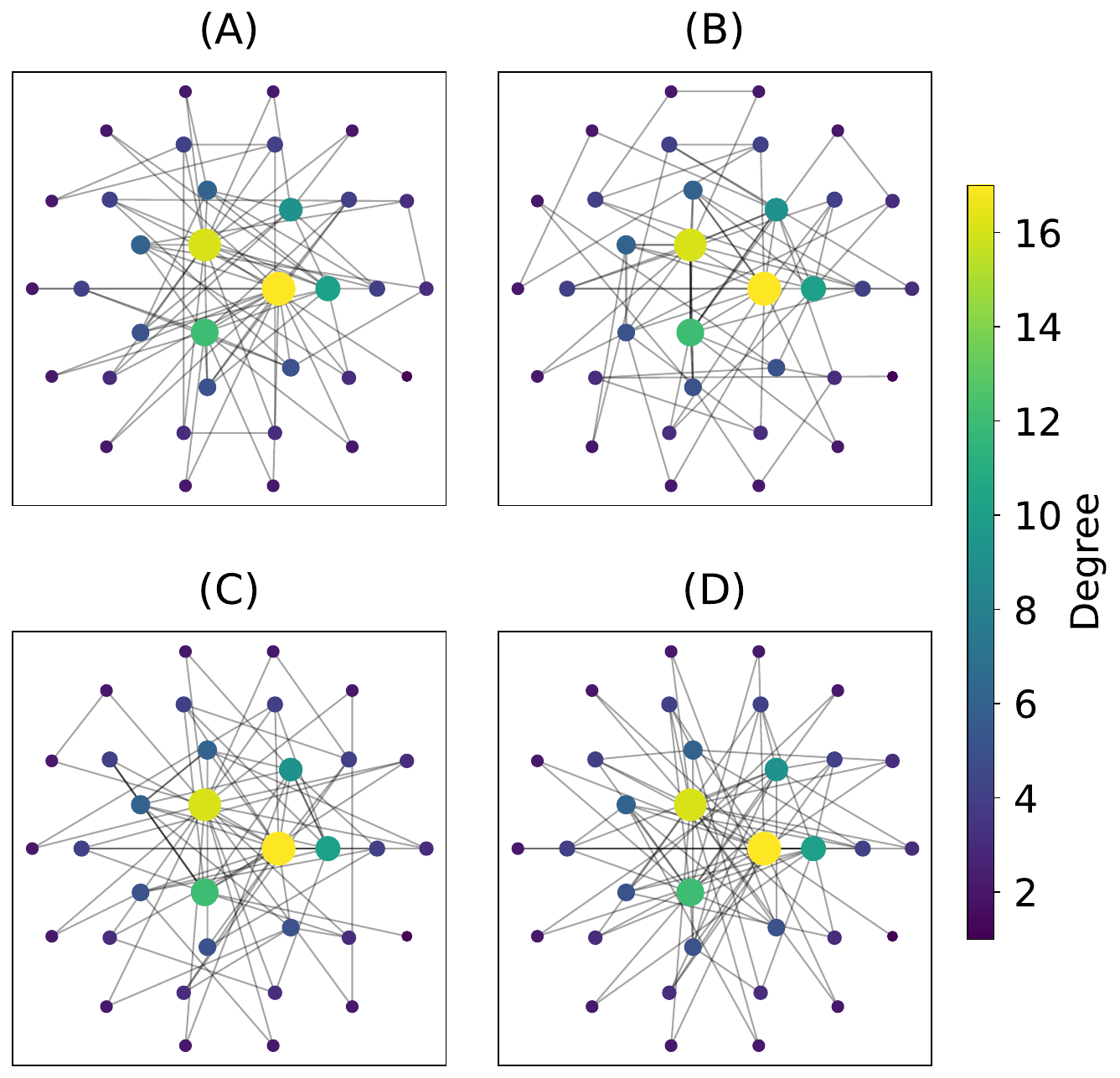}
\caption{(A) Visualization of the j karate club's network~\cite{zachary1977information}. Nodes' positions depend on their degree, with high-degree nodes placed closer to the center of the visualization, and low-degree nodes appearing in the periphery. (B-D) Same as in (A), but for the networks respectively obtained via the configuration model (CM), the degree-biased configuration model (DBCM) with $p=1$, and simulated annealing (SA) optimization. The average shortest-path length of the original graph is $\langle \ell \rangle_{\text{real}} = 2.41$; 
for the other methods, we measure $\langle \ell \rangle_{p=0} = 2.54 \pm 0.08$, $\langle \ell \rangle_{p=1} = 2.25 \pm 0.02$ and $\langle \ell \rangle_{SA} = 2.16 \pm 0.03$, respectively.}
\label{fig:karate_structure}
\end{figure}

To provide intuition on the geometrical structure obtained by the different optimization methods, in Figure~\ref{fig:karate_structure}, we show 
the network obtained by using as input the degree sequence of the Zachary karate club's network~\cite{zachary1977information}.
One can see that the network structure that minimize the average shortest-path length, obtained via SA optimization, has a star-like geometry. A relatively similar structure is also obtained via DBCM with $p=1$.

\section{Conclusion}
In this work, we investigated the problem of finding, given an input degree sequence, the network structure with the smallest average shortest-path length. This problem is particularly relevant in infrastructural networks devoted to transportation and communication, where the number of connections at each node is often limited by physical or economic constraints. In such systems, the overall efficiency of the system strongly depends on its global geometry.

While this problem admits an exact solution in the case of trees, using the greedy-tree (GT) construction algorithm, for general graphs with cycles, the search space grows exponentially and exact optimization becomes infeasible. In this paper, we proposed one scalable heuristic approach named degree-biased configuration model (DBCM), that combines the standard configuration model (CM) with the GT algorithm.

We performed systematic experiments on synthetic power-law degree sequences and found that the DBCM method consistently outperforms the standard CM, especially with degree sequences displaying intermediate levels of heterogeneity. We further tested DBCM performance on a corpus of 109 real-world networks. We found that DBCM generally yield networks that have an average shortest-path length significantly smaller than the actual real network structures. DBCM performance is also quite close to the one achieved via simulated annealing (SA) optimization, in spite DBCM is significantly faster than SA.

To conclude, our results highlight the impact of degree placement and hierarchical organization on the global efficiency of the network. We have shown that, even under strict degree constraints, there exists significant flexibility in how edges can be arranged, and that this flexibility can be exploited to systematically reduce pairwise distances. 

These findings open several directions for future research. On the methodological side, our DBCM algorithm could be extended to incorporate additional structural constraints, and generalized to weighted or directed networks. It would also be interesting to investigate alternative constructions, such as an edge-based version of our algorithm, as opposed to the current node-based one, and to analyze the trade-offs between minimizing global distances and preserving other properties such as robustness, modularity and congestion \cite{albert2000error,schneider2011mitigation}. From a geometric perspective, it would be interesting to study distance minimization in networks embedded in hyperbolic space \cite{krioukov2010hyperbolic, boguna2010sustaining}, or to interpret the optimized configurations in terms of degree-degree correlations and rich-club properties \cite{colizza2006detecting}, as we observe that minimizing path lengths is related to having connections among high-degree nodes. Finally, our framework may be relevant to communication systems where there are degree constraints, such as quantum communication and entanglement networks \cite{kimble2008quantum, meng2025path, radicchi2026enabling}. An interesting question to address would be characterizing how these optimized networks affect processes such as epidemic spreading or information diffusion \cite{pastor2015epidemic}, for instance in the context of social media campaigns. Similarly, exploring the spectral properties of the optimized graphs could provide further insights into their diffusion and transport efficiency \cite{van2023graph}. Moreover, machine-learning-based approaches could be investigated, for example one could train a neural network on solutions provided by simulated annealing on small networks, with the goal of generating accurate approximations for larger networks where direct optimization becomes computationally expensive \cite{grassia2021machine}.

\section{Appendix}\label{sec:appendix}
\subsection{Description of the DBCM algorithm}\label{sec:Implemented_algorithms}

Algorithm~\ref{alg:stochastic_greedy_conf} summarizes our implementation of the degree-biased configuration model (DBCM).


\begin{algorithm}[H]
\caption{Degree-Biased Configuration Model (DBCM)}
\label{alg:stochastic_greedy_conf}
\begin{algorithmic}[1]

\Input
   \Desc{The degree sequence
$\mathbf{k} = (k_1, \ldots, k_N)$} 
   \Desc{The  model parameter $p \in [0,1]$} 
  \EndInput
  \Output
  \Desc{A multigraph $G$}  
  \EndOutput

\State Initialize empty multigraph $G$ with vertices corresponding to the input degree sequence and parameter $p \in [0, 1]$
\State Initialize max-heap $H$ ordered by available degree $k_n^{(a)} = k_n$ for $n = 1, \ldots, N$
\State Find $r = \arg \max_{n \in  H} k^{(a)}_n$ and remove $r$ from $H$
\State Add $r$ to $Q$

\While{$Q \neq \emptyset$ and $H \neq \emptyset$}
    \State Remove $u$ from $Q$
    \While{
    $k^{(a)}_u > 0$
    and $H \neq \emptyset$}
        \State Extract vertex $v = \arg \max_{n \in  H} k^{(a)}_n$ 
        \If {random[0, 1] < p}
            \State  Add edge $(u,v)$ to $G$
            \State  Decrease 
            $k^{(a)}_u \to k^{(a)}_u -1$ and
            $k^{(a)}_v \to k^{(a)}_v -1$
            \If {
            $k^{(a)}_v > 0$
            }
                \State Add $v$ to $Q$
            \EndIf
        \EndIf
    \EndWhile
\EndWhile

\State Generate edges using the configuration model with the input sequence of available degrees $\mathbf{k^{(a)}}$

\State Add these edges to $G$

\State \Return $G$
\end{algorithmic}
\end{algorithm}

\subsection{Average shortest-path length sampling}\label{sec:appendix_avg_sht_path_sampling_k_roots}

To validate the accuracy of the sampling estimator, we compare it against the exact average shortest-path length on small networks, where the exact computation is feasible. For each $N \in \{250, 500, 750, 1000\}$, we generate 5 independent graph realizations using our degree-biased configuration model with $p=1$, generated from a power-law degree distribution with parameters $\lambda = 2.5$ and minimum degree 3.

For each graph, we compute the exact average shortest-path length and compare it to the sampled estimator using different number of sampled roots. In particular, we test $Z \in \{1,2,5,10,20,50,100\}$. For each value of $Z$, we repeat the sampling procedure with 5 independent sampling seeds. 
We then measure the relative error 
defined as 
\begin{equation}
    \epsilon = {\frac{|\langle \ell (G) \rangle - \overline{ \ell (G) }|}{\overline{ \ell (G) }}},
\end{equation}
where $\langle \ell (G) \rangle$ is the estimated average shortest-path length obtained by sampling $Z$ root nodes, and $\overline{ \ell (G) }$ is the exact average shortest-path length.

\begin{figure}[h]
    \centering
\includegraphics[width=\columnwidth]{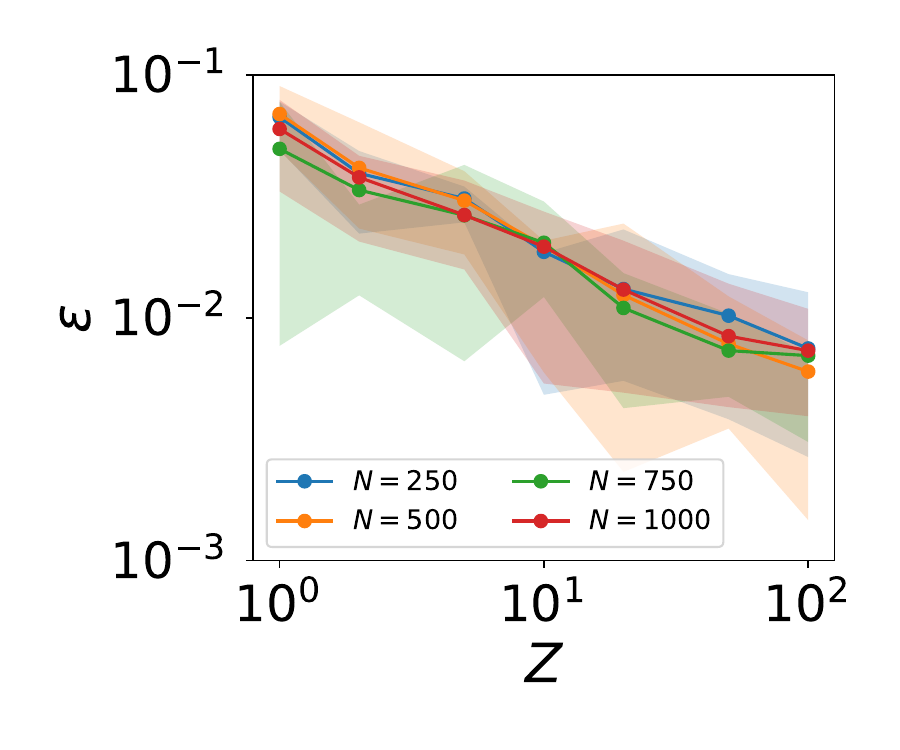}
\caption{Relative error $\epsilon$ of the sampled average shortest-path length estimator as a function of the number of sampled roots $Z$, shown on a log--log scale. Solid lines denote the mean relative error across all realizations, while the shaded regions indicate the interquartile range (IQR; 25th--75th percentiles). Results are obtained from 5 independent graph realizations of the DBCM with $p=1$, generated from a power-law degree distribution with exponent $\lambda=2.5$ and minimum degree $k_{\min}=3$, and 5 independent sampling seeds per graph. Different colors correspond to different network sizes $N$.}
\label{fig:sampling_validation}
\end{figure}

Figure~\ref{fig:sampling_validation} shows a clear convergence as $Z$ increases. In particular, for $Z = 20$, the relative error is below 2\% for all network sizes, which indicates that the sampling estimator provides a reliable approximation of the average shortest-path length.

\subsection{Empirical time complexity analysis of our degree-biased configuration model}\label{sec: appendix_time_complexity}

In this section we perform an empirical time complexity analysis of the DBCM graph construction and the sampling average shortest-path length estimator.

We generate 5 independent degree sequences from a power-law distribution with parameters $\lambda = 2.5$, $_{\min} = 3$, and network size $N$ ranging from $N = 5 \time 10^3$ to $N = 10^6$. For each $N$ and degree sequence, we construct graphs using the DBCM algorithm with $p = 1$ and measure the time required to generate each graph. We then compute the sampled average shortest-path length using $Z = 20$ source nodes and record the execution time. Finally, we average the runtimes corresponding to the graph construction, the sampling of the distances, and the total time.

\begin{figure}[tb]
    \centering
\includegraphics[width=\columnwidth]{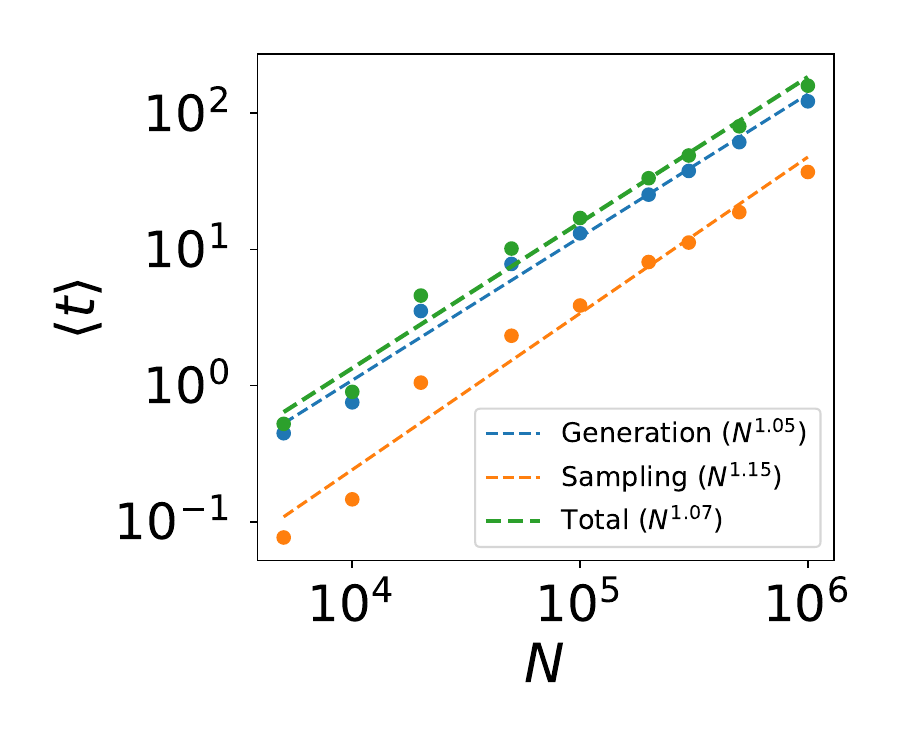}
\caption{Empirical time complexity analysis of the DBCM graph generation with $p=1$, in blue, and the distance sampling computation, in orange, for the average shortest-path length as a function of the network size. Total time is shown in green.
$\langle t \rangle$ denotes the average execution time, in seconds, averaged over 5 random seeds; and dashed lines correspond to power-law fits obtained by a linear regression in log-log scale.}
\label{fig:time_complexity}
\end{figure}

To estimate the scaling behavior, we fit power-law models of the form $T(N) \sim N^{\alpha}$ in log-log scale and represent them in Figure~\ref{fig:time_complexity}. We obtain empirical exponents $\alpha = 1.05$ for the graph construction and $\alpha = 1.15$ for the computation of the sampled average shortest-path length. The total runtime scales with the exponent $\alpha = 1.07$. These values indicate a scaling close to linear over the tested range of network sizes.

\subsection{$p$ parameter selection plots and extended results for $k_{\min} = 5$ and varying $\lambda$}\label{sec:appendix_p_selection}

\begin{figure*}[t]
    \centering
\includegraphics[width=\textwidth]{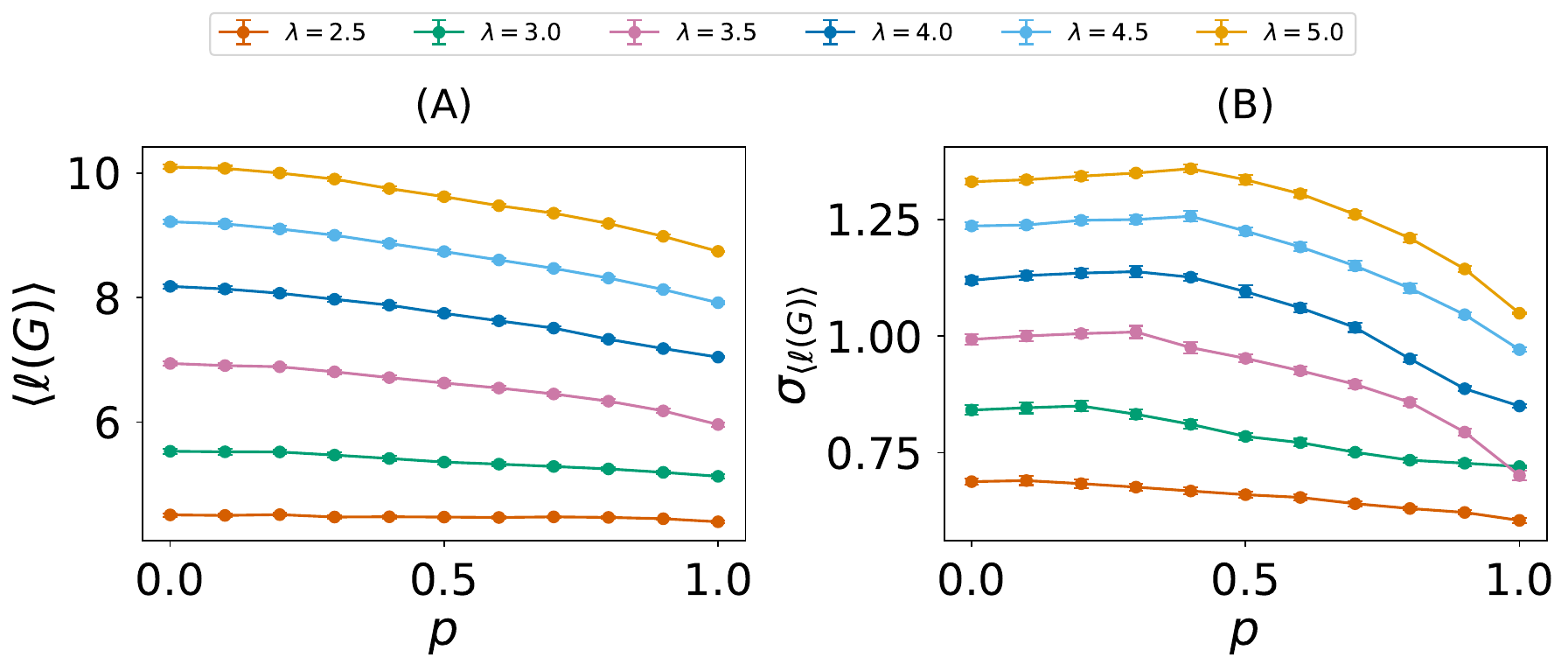}
\caption{(A) Average shortest-path length $\langle \ell (G) \rangle$ and (B) standard deviation of shortest path distances $\sigma_{\langle \ell (G) \rangle}$ as functions of the stochastic parameter $p$, for networks of size $N=10^5$ with minimum degree $k_{\min}=3$. Results are averaged over 20 independent degree sequences, with 10 graph realizations per sequence and per value of $p$. Error bars represent the standard deviation across degree sequences. 
}
\label{fig:p_parameter_selection}
\end{figure*}

\begin{figure}[t]
   \centering
\includegraphics[width=\columnwidth]{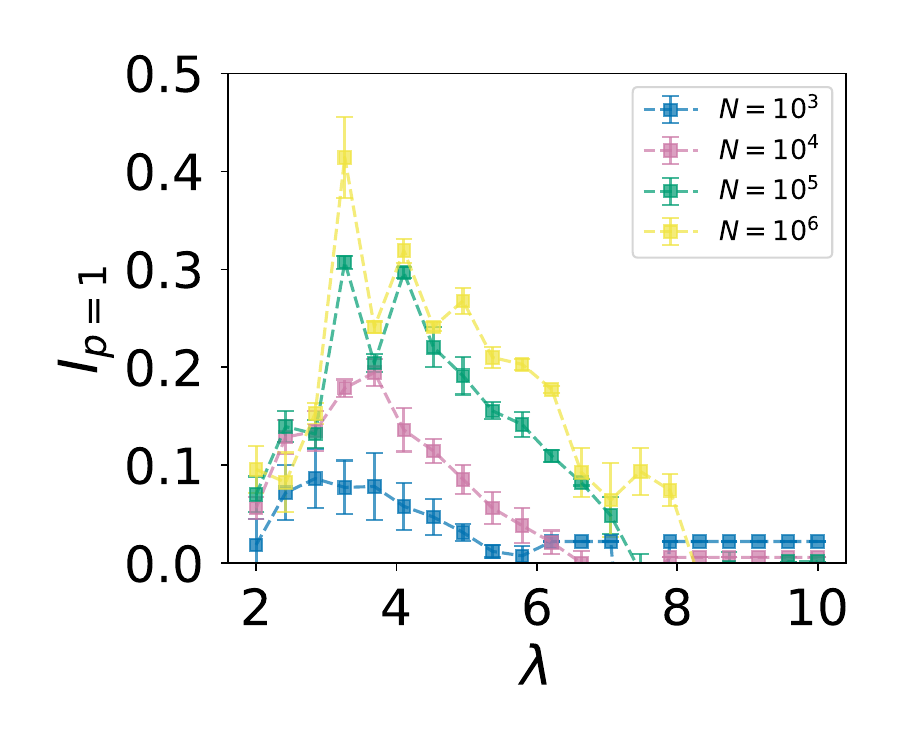}
\caption{Standard deviation ($\sigma$) 
of the relative improvement of the DBCM construction with $p=1$ over $p=0$ as a function of the power-law exponent $\lambda$, for networks of varying sizes $n$ and minimum degree $k_{\min}=3$. Results are averaged over 50 independent degree sequences for each $\lambda$ and 50 independent graph realizations of parameter $p$ for each degree sequence. 
}
\label{fig:lambda_rel_improvement_xm3_std}
\end{figure}

\begin{figure*}[!htb]
    \centering
\includegraphics[width=\textwidth]{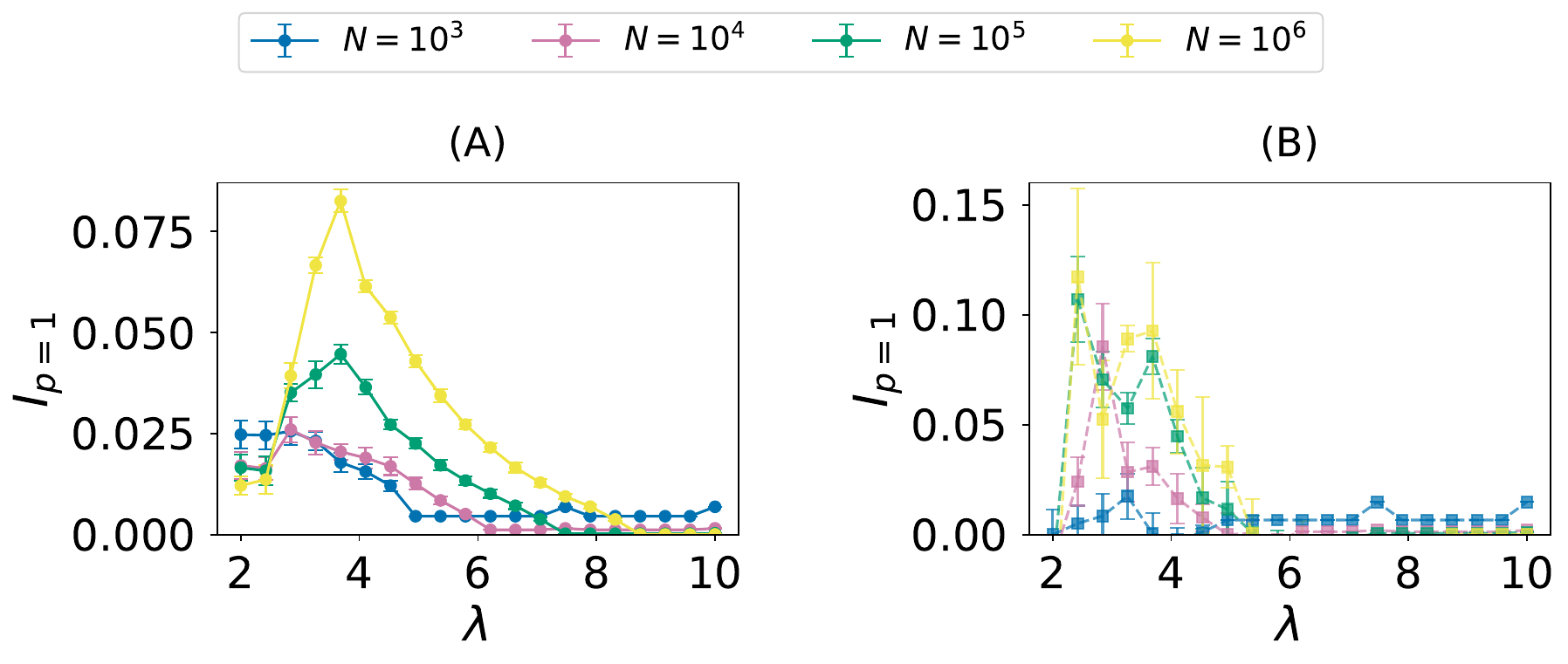}
\caption{(A) Mean and (B) standard deviation of the relative improvement of the DBCM construction with $p=1$ over $p=0$ as a function of the power-law exponent $\lambda$, for networks of varying sizes $n$ and minimum degree $x_m=5$. Results are averaged over 30 independent degree sequences for each $\lambda$ and 30 independent graph realizations of parameter $p$ for each degree sequence. 
}
\label{fig:lambda_rel_improvement_xm5}
\end{figure*}

We provide complementary plots to Figure~\ref{fig:lambda_rel_improvement} supporting that $p=1$ yields the lowest average shortest-path length among all $p$ values. In particular, in Figure~\ref{fig:p_parameter_selection} we plot the average shortest-path length and its standard deviation as functions of $p$, for different values of $\lambda$. We observe a pronounced decrease in both quantities as $p$ is increased. Moreover, in Figure~\ref{fig:lambda_rel_improvement_xm3_std} we plot the standard deviation of the relative improvement of the DBCM construction with $p=1$ over $p=0$ as a function of the power-law exponent $\lambda$, for networks of minimum degree $x_m=3$, and in Figure~\ref{fig:lambda_rel_improvement_xm5} we plot the mean and standard deviation, for networks of minimum degree $x_m=5$.

\subsection{Distance distributions from synthetic power-law networks}\label{subsec_app: histograms}

\begin{figure*}[t]
    \centering
\includegraphics[width=\textwidth]{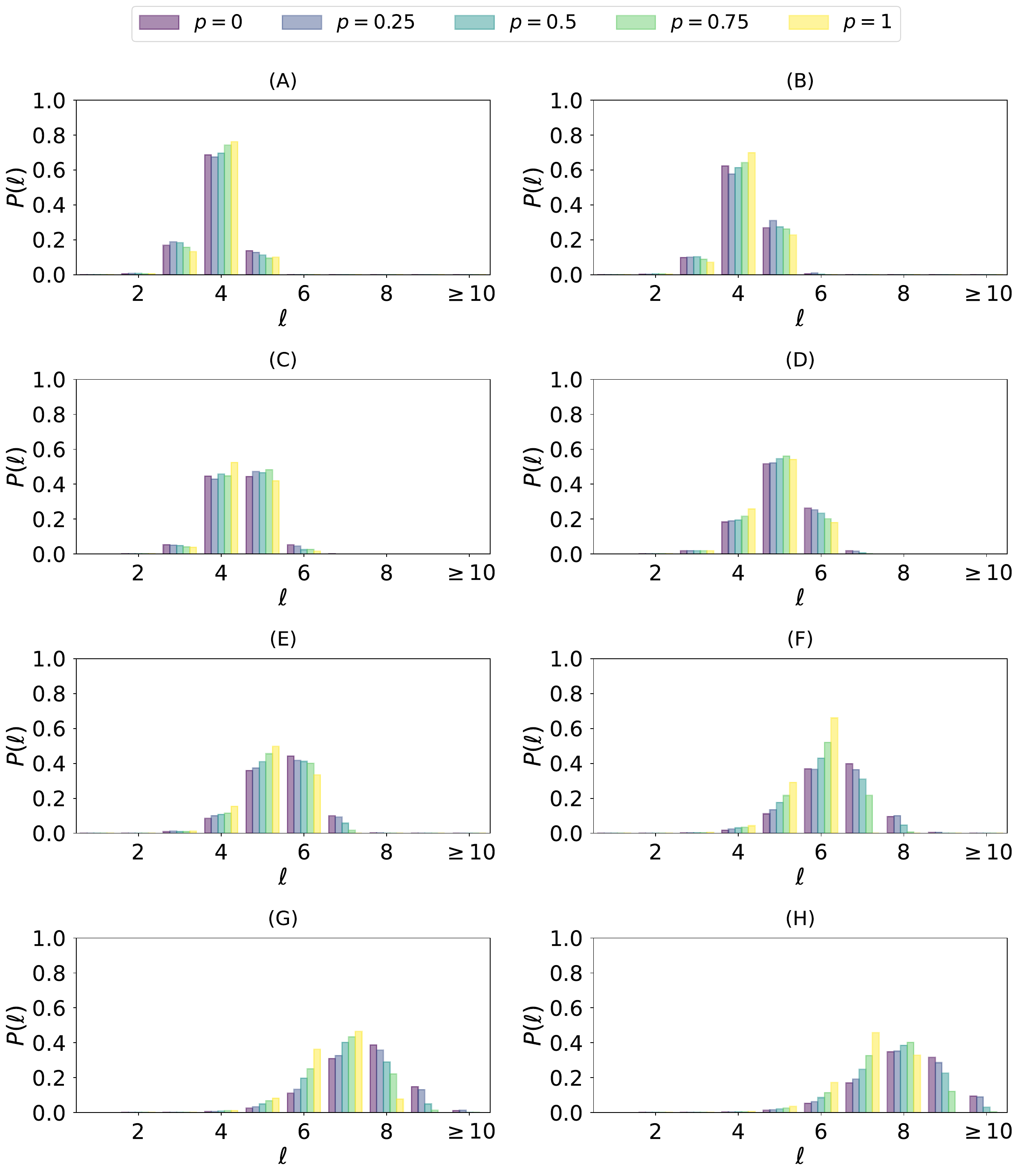}
\caption{Shortest-path length distributions for power-law degree sequences with $N=10^5$ and varying exponent $\lambda \in \{2.1, 2.3, 2.5, 2.8, 3.0, 3.3, 3.7, 4.0 \}$, corresponding to panels (A) to (H), respectively. For each $\lambda$, distances are computed for five realizations of degree sequences and shown for different values of $p$. Bars represent the fraction of sampled node pairs $P(\ell)$ at each distance $\ell$. Mean and standard deviation for each $p$ are reported in Table~\ref{tab:power_law_histograms_stats}.}
\label{fig:power_law_histograms}
\end{figure*}

In Figure~\ref{fig:power_law_histograms} we report the sampled shortest-path length distributions for power-law degree sequences with exponent $\lambda \in \{2.1, 2.3, 2.5, 2.8, 3.0, 3.5, 4.0\}$ and network size $N=10^5$. For each value of $\lambda$, we generate five independent degree sequences and, for each sequence, we construct graphs using the DBCM with $p \in \{0, 0.25, 0.5, 0.75, 1\}$. We estimate the shortest path distances by sampling 20 root nodes per graph. For each $\lambda$ we show a histogram of the fraction of sampled node pairs within each bin, where each bin corresponds to a shortest-path length, with the last bin containing distances greater than or equal to 10. We also report, for each $p$, the average shortest-path length and the standard deviation of distances.

\begin{table*}[h]
\centering
\label{tab:power_law_histograms_stats}
\setlength{\tabcolsep}{14pt}
\begin{tabular}{cccccc}
\hline
$\lambda$ & $p=0.00$ & $p=0.25$ & $p=0.50$ & $p=0.75$ & $p=1.00$ \\
\hline
$2.1$ & $3.96\pm0.58$ & $3.93\pm0.59$ & $3.92\pm0.56$ & $3.92\pm0.52$ & $3.95\pm0.51$ \\
$2.3$ & $4.18\pm0.61$ & $4.22\pm0.64$ & $4.17\pm0.62$ & $4.17\pm0.59$ & $4.15\pm0.54$ \\
$2.5$ & $4.49\pm0.69$ & $4.51\pm0.68$ & $4.46\pm0.64$ & $4.49\pm0.63$ & $4.41\pm0.60$ \\
$2.8$ & $5.07\pm0.77$ & $5.05\pm0.77$ & $5.01\pm0.73$ & $4.95\pm0.71$ & $4.88\pm0.71$ \\
$3.0$ & $5.54\pm0.84$ & $5.48\pm0.86$ & $5.40\pm0.81$ & $5.30\pm0.74$ & $5.15\pm0.72$ \\
$3.3$ & $6.45\pm0.91$ & $6.39\pm0.97$ & $6.16\pm0.91$ & $5.94\pm0.80$ & $5.61\pm0.60$ \\
$3.7$ & $7.54\pm1.04$ & $7.44\pm1.07$ & $7.06\pm1.00$ & $6.82\pm0.95$ & $6.51\pm0.81$ \\
$4.0$ & $8.18\pm1.13$ & $8.10\pm1.15$ & $7.78\pm1.08$ & $7.48\pm0.99$ & $7.06\pm0.84$ \\
\hline
\end{tabular}
\caption{Shortest-path length distribution statistics for power-law degree sequences with $N=10^5$ and varying exponent $\lambda \in \{2.1, 2.3, 2.5, 2.8, 3.0, 3.3, 3.7, 4.0 \}$. For each $\lambda$, we report the mean $\pm$ standard deviation, computed for five realizations of degree sequences and different values of $p$ (see Figure~\ref{fig:power_law_histograms}).}
\end{table*}

From the histograms in Figure~\ref{fig:power_law_histograms} we observe that as the power-law exponent $\lambda$ increases and the degree distribution becomes more homogeneous, the distribution shifts to larger distances and becomes wider. For small $\lambda$ values, most node pairs are at distance 3 or 4 with each other, due to the presence of hubs. While for larger $\lambda$ values, distances spread over a broader range and both the mean and standard deviation increase. Moreover, for each fixed $\lambda$, increasing $p$ shifts the distribution toward shorter distances. In particular, the average shortest-path length decreases as $p$ increases, and the tail of larger distances becomes less pronounced.

\subsection{Assortativity analysis of synthetic power-law networks}\label{subsec_appendix: assortativity}

\begin{figure*}[t]
    \centering
\includegraphics[width=\textwidth]{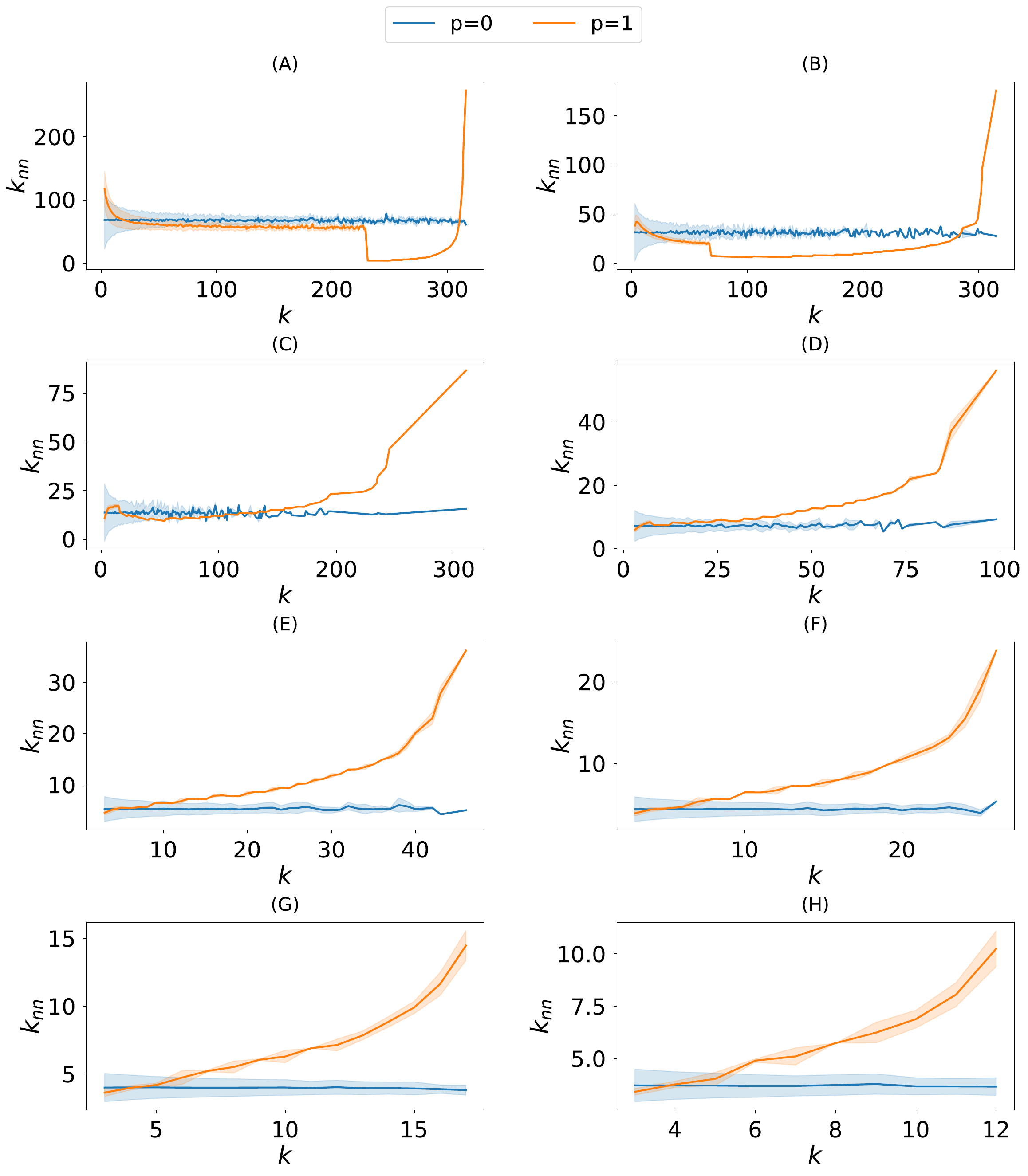}
\caption{Average degree of neighbors $k_{nn}$ as a function of node degree $k$ for synthetic power-law networks generated with different $\lambda$ exponents and $N=10^5$. Each panel (A) to (H) corresponds to one representative realization of $\lambda \in \{3.0, 3.5, 4.0, 4.5, 5.0, 5.5, 6.0, 6.5\}$. Curves show the mean average degree of neighbors over all nodes with the same degree, for the $p=0$ and $p=1$ constructions. Shaded regions represent one standard deviation across nodes with the same degree.} 
\label{fig:assortativity_plots_node_degree}
\end{figure*}

\begin{figure*}[t]
    \centering
\includegraphics[width=\textwidth]{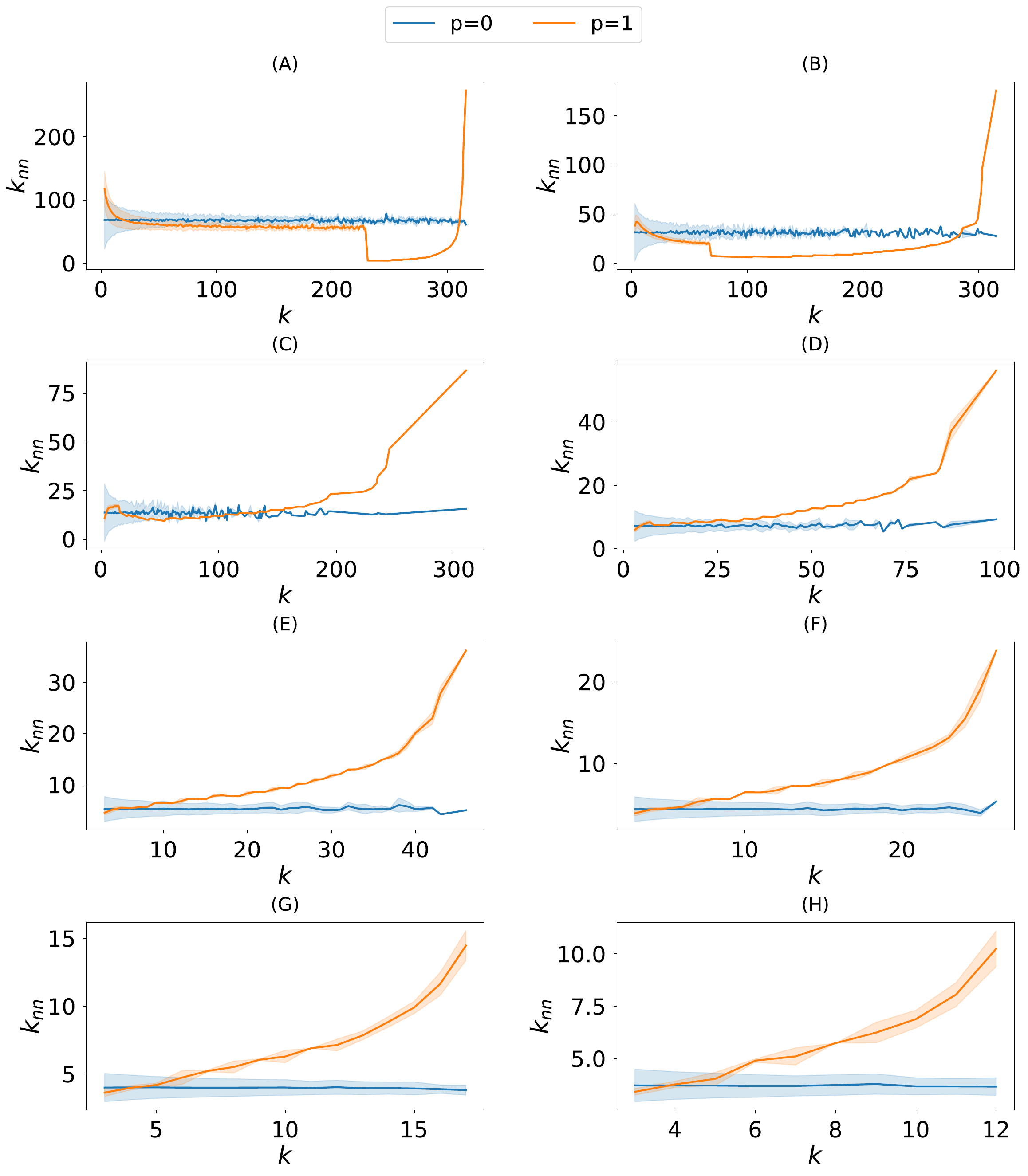}
\caption{Evolution of the graph average neighbor degree $k_{nn}$ as a function of the number of edges $E$ during the network construction process for synthetic power-law networks with different $\lambda$ exponents and $N = 10^5$. Each panel (A) to (H) corresponds to one value of $\lambda \in \{3.0, 3.5, 4.0, 4.5, 5.0, 5.5, 6.0, 6.5\}$. The horizontal axis represents the cumulative number of edges added during graph construction, while the vertical axis shows the instantaneous average degree of neighbors over all nodes in the partially constructed graph. Curves are shown for both the $p=0$ and $p=1$ constructions using a representative realization.}
\label{fig:assortativity_plots_num_edges}
\end{figure*}

Figures~\ref{fig:assortativity_plots_node_degree} and \ref{fig:assortativity_plots_num_edges} examine the degree correlations generated by the DBCM procedure for synthetic power-law degree sequences with different $\lambda$ exponents.

Figure~\ref{fig:assortativity_plots_node_degree} shows the average degree of neighbors as a function of node degree, $\langle k_{nn}(k)\rangle$, for networks generated with $p=0$ and $p=1$. For $p=0$ the curves remain approximately flat across all node degrees, consistent with the absence of degree correlations expected in the configuration model. In contrast, the greedy construction ($p=1$) produces a strong increase of $\langle k_{nn}(k)\rangle$ with node degree. This effect is strongest for medium values of $\lambda$, consistent with previous observations (see Figure~\ref{fig:lambda_rel_improvement}). 

Figure~\ref{fig:assortativity_plots_num_edges} illustrates the temporal evolution of the graph average neighbor degree as edges are added to the graph. In the $p=1$ construction, the average neighbor degree initially increases rapidly as high-degree nodes become interconnected early in the process. After reaching a maximum, corresponding to the formation of the tree backbone, the quantity decreases toward its final value as lower-degree nodes are progressively attached to the already formed hub core. By contrast, the $p=0$ construction evolves more smoothly and monotonically, reflecting the absence of preferential structural organization.

\begin{figure}[t]
    \centering
\includegraphics[width=\columnwidth]{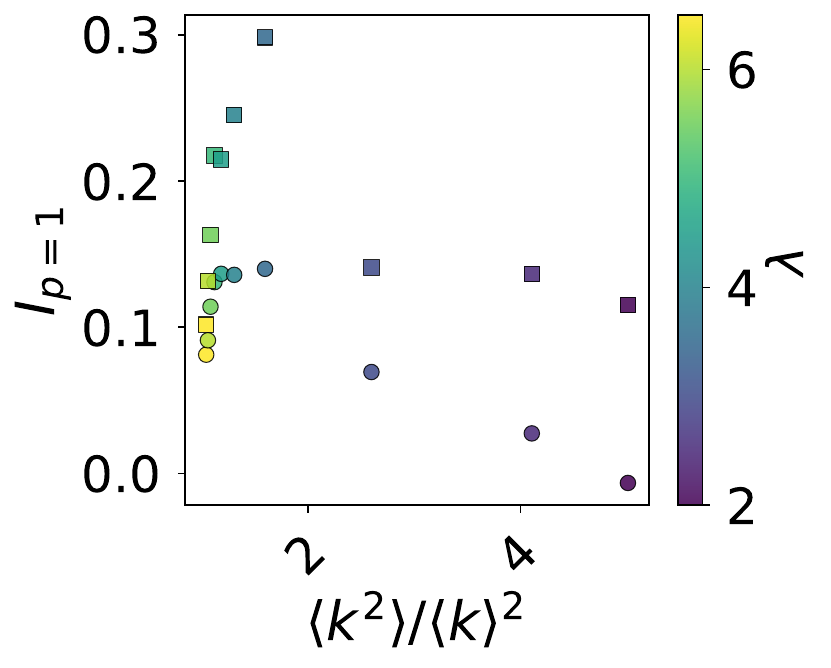}
\caption{Relative improvement obtained by the $p=1$ construction over the $p=0$ construction as a function of the degree heterogeneity $\langle k^2 \rangle /\langle k \rangle ^2$. Circles represent improvement in the mean shortest-path length, while squares represent improvement in the standard deviation of shortest-path lengths. Each point corresponds to one value of the power-law exponent $\lambda$, with colors indicating its value.}
\label{fig:assortativity_plot_deg_heterogeneity}
\end{figure} 

\subsection{Real networks description}\label{sec: appendix_table_real_networks}
Table~\ref{tab:real_networks} summarizes the 109 real-world networks used in our experiments. We report their size, a category to which they belong, and bibliographic references.

\begin{table*}[t]
\centering
\footnotesize
\setlength{\tabcolsep}{2pt}
\begin{tabular}{lrrr @{\hspace{1.2cm}} lrrr}
\toprule
Network & $N$ & $M$ & Ref. & Network & $N$ & $M$ & Ref. \\
\midrule
Social 3 {\color{gray}(Soc)} & 32 & 80 & \cite{milo2004superfamilies} & DBLP, citations {\color{gray}(Cit)} & 12591 & 49620 & \cite{ley2002dblp,konect} \\
Karate club {\color{gray}(Soc)} & 34 & 78 & \cite{zachary1977information} & Spanish (Lasagne) {\color{gray}(Info)} & 12643 & 55019 & \cite{konect} \\
Protein 2 {\color{gray}(Bio)} & 53 & 123 & \cite{milo2004superfamilies} & Google {\color{gray}(Info)} & 15763 & 148585 & \cite{palla2007directed} \\
Dolphins {\color{gray}(Soc)} & 62 & 159 & \cite{lusseau2003bottlenose} & Astrophysics {\color{gray}(Cit)} & 16046 & 121251 & \cite{newman2001structure} \\
Social 1 {\color{gray}(Soc)} & 67 & 142 & \cite{milo2004superfamilies} & Cond-Mat, 1995-1999 {\color{gray}(Cit)} & 16264 & 47594 & \cite{newman2001structure} \\
Les Miserables {\color{gray}(Soc)} & 77 & 254 & \cite{knuth1993stanford} & AstroPhys, 1993-2003 {\color{gray}(Cit)} & 18771 & 198050 & \cite{leskovec2007graph} \\
Protein 1 {\color{gray}(Bio)} & 95 & 213 & \cite{milo2004superfamilies} & Gnutella, Aug. 25, 2002 {\color{gray}(Tech)} & 22687 & 54705 & \cite{ripeanu2002mapping,leskovec2007graph} \\
Protein 3 {\color{gray}(Bio)} & 97 & 212 & \cite{milo2004superfamilies} & Internet {\color{gray}(Tech)} & 22963 & 48436 & \cite{newman_internet_2006} \\
Political books {\color{gray}(Soc)} & 105 & 441 & \cite{adamic2005political} & Thesaurus {\color{gray}(Info)} & 23132 & 297094 & \cite{kiss1973associative,konect} \\
David Copperfield {\color{gray}(Soc)} & 112 & 425 & \cite{newman2006finding} & Cond-Mat, 1993-2003 {\color{gray}(Cit)} & 23133 & 93439 & \cite{leskovec2007graph} \\
College football {\color{gray}(Soc)} & 115 & 613 & \cite{girvan2002community} & Cora {\color{gray}(Cit)} & 23166 & 89157 & \cite{vsubelj2013model,konect} \\
S 208 {\color{gray}(Soc)} & 122 & 189 & \cite{milo2004superfamilies} & AS Caida {\color{gray}(Tech)} & 26475 & 53381 & \cite{leskovec2005graphs} \\
High school, 2011 {\color{gray}(Soc)} & 126 & 1709 & \cite{fournet2014contact} & Gnutella, Aug. 24, 2002 {\color{gray}(Tech)} & 26518 & 65369 & \cite{ripeanu2002mapping,leskovec2007graph} \\
Bay Wet {\color{gray}(Bio)} & 128 & 2075 & \cite{konect} & Linux, mailing list {\color{gray}(Comm)} & 26885 & 159996 & \cite{konect} \\
Bay Dry {\color{gray}(Bio)} & 128 & 2106 & \cite{ulanowicz1998network,konect} & Hep-Th, citations {\color{gray}(Cit)} & 27769 & 352285 & \cite{leskovec2007graph,konect} \\
Radoslaw Email {\color{gray}(Comm)} & 167 & 3250 & \cite{radoslaw,konect} & Digg {\color{gray}(Soc)} & 30360 & 85155 & \cite{de2009social,konect} \\
High school, 2012 {\color{gray}(Soc)} & 180 & 2220 & \cite{fournet2014contact} & Cond-Mat, 1995-2003 {\color{gray}(Cit)} & 30460 & 120029 & \cite{newman2001structure} \\
Little Rock Lake {\color{gray}(Bio)} & 183 & 2434 & \cite{martinez1991artifacts,konect} & Linux, soft. {\color{gray}(Tech)} & 30834 & 213217 & \cite{konect} \\
Jazz {\color{gray}(Soc)} & 198 & 2742 & \cite{gleiser2003community} & Hep-Ph, citations {\color{gray}(Cit)} & 34546 & 420877 & \cite{leskovec2007graph,konect} \\
S 420 {\color{gray}(Soc)} & 252 & 399 & \cite{milo2004superfamilies} & Gnutella, Aug. 30, 2002 {\color{gray}(Tech)} & 36682 & 88328 & \cite{ripeanu2002mapping,leskovec2007graph} \\
C. Elegans, neural {\color{gray}(Bio)} & 297 & 2148 & \cite{watts1998collective} & Enron {\color{gray}(Comm)} & 36692 & 183831 & \cite{leskovec2009community} \\
Dublin {\color{gray}(Infra)} & 410 & 2765 & \cite{isella2011s,konect} & Cond-Mat, 1995-2005 {\color{gray}(Cit)} & 39577 & 175692 & \cite{newman2001structure} \\
US Air Trasportation {\color{gray}(Trans)} & 500 & 2980 & \cite{colizza2007reaction} & Slashdot {\color{gray}(Comm)} & 51083 & 116573 & \cite{gomez2008statistical,konect} \\
S 838 {\color{gray}(Soc)} & 512 & 819 & \cite{milo2004superfamilies} & Gnutella, Aug. 31, 2002 {\color{gray}(Tech)} & 62586 & 147892 & \cite{ripeanu2002mapping,leskovec2007graph} \\
Yeast, transcription {\color{gray}(Bio)} & 688 & 1078 & \cite{milo2002network} & Facebook {\color{gray}(Soc)} & 63731 & 817090 & \cite{viswanath2009evolution} \\
URV email {\color{gray}(Comm)} & 1133 & 5451 & \cite{guimera2003self} & Epinions {\color{gray}(Soc)} & 75879 & 405740 & \cite{richardson2003trust,konect} \\
Political blogs {\color{gray}(Soc)} & 1224 & 16715 & \cite{adamic2005political} & Slashdot zoo {\color{gray}(Soc)} & 79116 & 467731 & \cite{kunegis2009slashdot,konect} \\
Air traffic {\color{gray}(Trans)} & 1226 & 2408 & \cite{konect} & Flickr {\color{gray}(Soc)} & 105938 & 2316948 & \cite{McAuley2012,konect} \\
Network Science {\color{gray}(Cit)} & 1461 & 2742 & \cite{newman2006finding} & Wikipedia, edits {\color{gray}(Info)} & 116836 & 2027871 & \cite{brandes2010structural,konect} \\
Yeast protein (Jeong) {\color{gray}(Bio)} & 1846 & 2203 & \cite{jeong2001lethality} & Petster, cats {\color{gray}(Soc)} & 149684 & 5448197 & \cite{konect} \\
Petster, hamster {\color{gray}(Soc)} & 1858 & 12534 & \cite{konect} & Gowalla {\color{gray}(Soc)} & 196591 & 950327 & \cite{cho2011friendship,konect} \\
UC Irvine {\color{gray}(Soc)} & 1899 & 13838 & \cite{opsahl2009clustering,konect} & Libimseti {\color{gray}(Info)} & 220970 & 17233144 & \cite{brozovsky2007recommender,kunegis2012online,konect} \\
Yeast protein (Bu) {\color{gray}(Bio)} & 2284 & 6646 & \cite{bu2003topological} & Amazon, Mar. 2, 2003 {\color{gray}(Info)} & 262111 & 899792 & \cite{leskovec2007dynamics} \\
Japanese {\color{gray}(Info)} & 2704 & 7998 & \cite{milo2004superfamilies} & EU email {\color{gray}(Comm)} & 265009 & 364481 & \cite{leskovec2007graph,konect} \\
Open flights {\color{gray}(Trans)} & 2939 & 15677 & \cite{opsahl2010node,konect} & Web Stanford {\color{gray}(Info)} & 281903 & 1992636 & \cite{leskovec2009community} \\
Tennis {\color{gray}(Soc)} & 4342 & 81867 & \cite{radicchi2011best} & DBLP, collaborations {\color{gray}(Cit)} & 317080 & 1049866 & \cite{ley2002dblp,konect} \\
US Power grid {\color{gray}(Infra)} & 4941 & 6594 & \cite{watts1998collective} & Web Notre Dame {\color{gray}(Info)} & 325729 & 1090108 & \cite{albert1999internet} \\
GR-QC, 1993-2003 {\color{gray}(Cit)} & 5241 & 14484 & \cite{leskovec2007graph} & Actor coll. net. {\color{gray}(Soc)} & 382219 & 15038083 & \cite{barabasi1999emergence,konect} \\
HT09 {\color{gray}(Comm)} & 5352 & 18481 & \cite{isella2011s} & CiteSeer {\color{gray}(Cit)} & 384054 & 1736145 & \cite{bollacker1998citeseer,konect} \\
Jung {\color{gray}(Tech)} & 6120 & 50290 & \cite{vsubelj2012software,konect} & MathSciNet {\color{gray}(Cit)} & 391529 & 873775 & \cite{palla2008fundamental} \\
Reactome {\color{gray}(Bio)} & 6229 & 146160 & \cite{joshi2005reactome,konect} & Amazon, Mar. 12, 2003 {\color{gray}(Info)} & 400727 & 2349869 & \cite{leskovec2007dynamics} \\
Gnutella, Aug. 8, 2002 {\color{gray}(Tech)} & 6301 & 20777 & \cite{ripeanu2002mapping,leskovec2007graph} & Amazon, Jun. 6, 2003 {\color{gray}(Info)} & 403394 & 2443408 & \cite{leskovec2007dynamics} \\
JDK {\color{gray}(Tech)} & 6434 & 53658 & \cite{konect} & Amazon, May 5, 2003 {\color{gray}(Info)} & 410236 & 2439437 & \cite{leskovec2007dynamics} \\
AS Oregon {\color{gray}(Tech)} & 6474 & 12572 & \cite{leskovec2005graphs} & Zhishi {\color{gray}(Info)} & 415624 & 2374044 & \cite{niu2011zhishi,konect} \\
English {\color{gray}(Info)} & 7381 & 44207 & \cite{milo2004superfamilies} & Petster, dogs {\color{gray}(Soc)} & 426816 & 8543549 & \cite{konect} \\
Hep-Th, 1995-1999 {\color{gray}(Cit)} & 7610 & 15751 & \cite{newman2001structure} & Road network PA {\color{gray}(Infra)} & 1088092 & 1541898 & \cite{leskovec2009community} \\
Gnutella, Aug. 9, 2002 {\color{gray}(Tech)} & 8114 & 26013 & \cite{ripeanu2002mapping,leskovec2007graph} & YouTube friend. net. {\color{gray}(Soc)} & 1134890 & 2987624 & \cite{leskovec2012,konect} \\
French {\color{gray}(Info)} & 8325 & 23841 & \cite{milo2004superfamilies} & Road network TX {\color{gray}(Infra)} & 1379917 & 1921660 & \cite{leskovec2009community} \\
Gnutella, Aug. 6, 2002 {\color{gray}(Tech)} & 8717 & 31525 & \cite{ripeanu2002mapping,leskovec2007graph} & AS Skitter {\color{gray}(Tech)} & 1696415 & 11095298 & \cite{leskovec2005graphs} \\
Gnutella, Aug. 5, 2002 {\color{gray}(Tech)} & 8846 & 31839 & \cite{ripeanu2002mapping,leskovec2007graph} & Road network CA {\color{gray}(Infra)} & 1965206 & 2766607 & \cite{leskovec2009community} \\
Hep-Th, 1993-2003 {\color{gray}(Cit)} & 9875 & 25973 & \cite{leskovec2007graph} & Wikipedia, pages {\color{gray}(Info)} & 2070486 & 42336692 & \cite{palla2008fundamental} \\
PGP {\color{gray}(Comm)} & 10680 & 24316 & \cite{boguna2004models} & US Patents {\color{gray}(Cit)} & 3774768 & 16518947 & \cite{hall2001nber,konect} \\
Gnutella, August 4 2002 {\color{gray}(Tech)} & 10876 & 39994 & \cite{ripeanu2002mapping,leskovec2007graph} & DBpedia {\color{gray}(Info)} & 3966895 & 12610982 & \cite{auer2007dbpedia,konect} \\
Spanish (book) {\color{gray}(Info)} & 11586 & 43065 & \cite{milo2004superfamilies} & LiveJournal {\color{gray}(Soc)} & 5203764 & 48709773 & \cite{mislove2007measurement,konect} \\
Hep-Ph, 1993-2003 {\color{gray}(Cit)} & 12006 & 118489 & \cite{leskovec2007graph} & & & & \\
\bottomrule
\end{tabular}
\caption{Summary of the real-world networks used in the experiments. For each network we report the number of nodes $n$, the number of edges $m$, and the corresponding reference. Network categories are indicated in parentheses: \textcolor{gray}{Bio} (biological), \textcolor{gray}{Soc} (social), \textcolor{gray}{Comm} (communication), \textcolor{gray}{Cit} (citation), \textcolor{gray}{Info} (information), \textcolor{gray}{Tech} (technological), \textcolor{gray}{Infra} (infrastructure), and \textcolor{gray}{Trans} (transportation).}
\label{tab:real_networks}
\end{table*}

\subsection{Connectivity probability of the DBCM construction with real networks}\label{appendix: connectivity_prob}

\begin{figure}[tb]
    \centering
\includegraphics[width=\columnwidth]{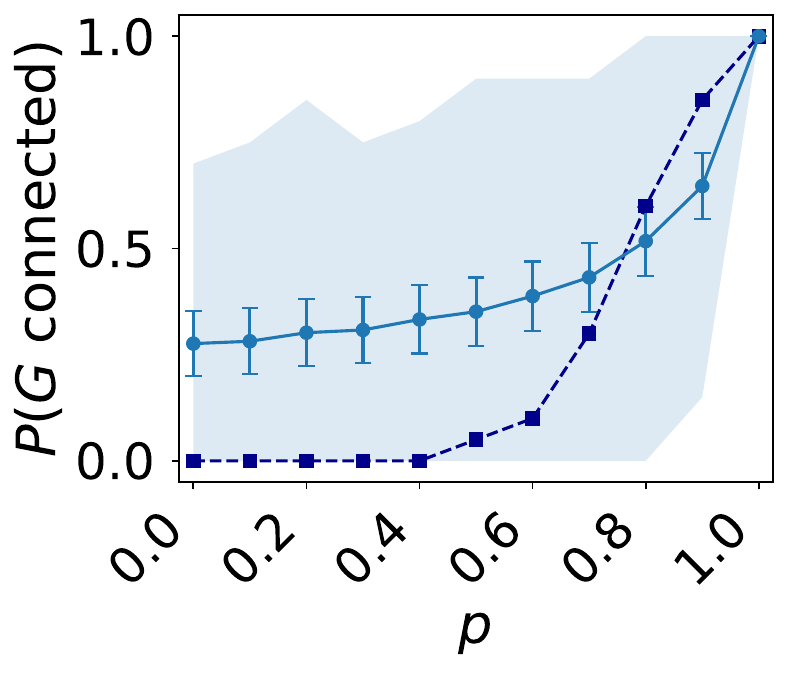}
\caption{Connectivity probability of the DBCM construction as a function of $p$. For each value of $p$, circles denote the mean probability of obtaining a connected graph across networks, averaged over 20 constructions, with 95\% confidence intervals (CI) denoted by the corresponding error bars. Squares in the dashed line indicate the median, while the shaded region represents the interquartile range (IQR). 
}
\label{fig:connectivity_probability}
\end{figure}

We analyze the probability that the DBCM construction yields a connected graph as a function of $p$. For each network and each value of $p$, we estimate the connectivity probability as the fraction of connected realizations over the 20 runs. Figure~\ref{fig:connectivity_probability} shows the mean connectivity probability across networks, together with the interquartile range (IQR, 25th-75th percentiles) and 95\% confidence intervals (CI).  The CI are computed as
$x \pm \frac{s}{\sqrt{N}}$,
where $x$ denotes the sample mean of the relative improvements, $s$ is the sample standard deviation across networks, and $N$ is the number of networks. We observe a strong dependence on $p$. For small values of $p$, the probability of obtaining a connected graph is significantly lower, while it increases rapidly as $p$ approaches 1, where connectivity is always guaranteed by construction. These results highlight that connectivity is a critical factor when evaluating the performance of the model, particularly for small values of $p$.

\subsection{Extended results on small real-world networks}\label{appendix: small real networks}

In this appendix, we provide additional results for the comparison between the DBCM construction and simulated annealing (SA) on small real-world networks.



To gain intuition on the structural differences created by the different network constructions, in Figure~\ref{fig:karate_structure} we show a realization of the Karate network \cite{zachary1977information} for each method: DBCM with $p=0$ and $p=1$, simulated annealing (SA) and the original real network. One can see that shorter average path lengths are related to a star-like shape.

\begin{figure}[t]
    \centering
\includegraphics[width=\columnwidth]{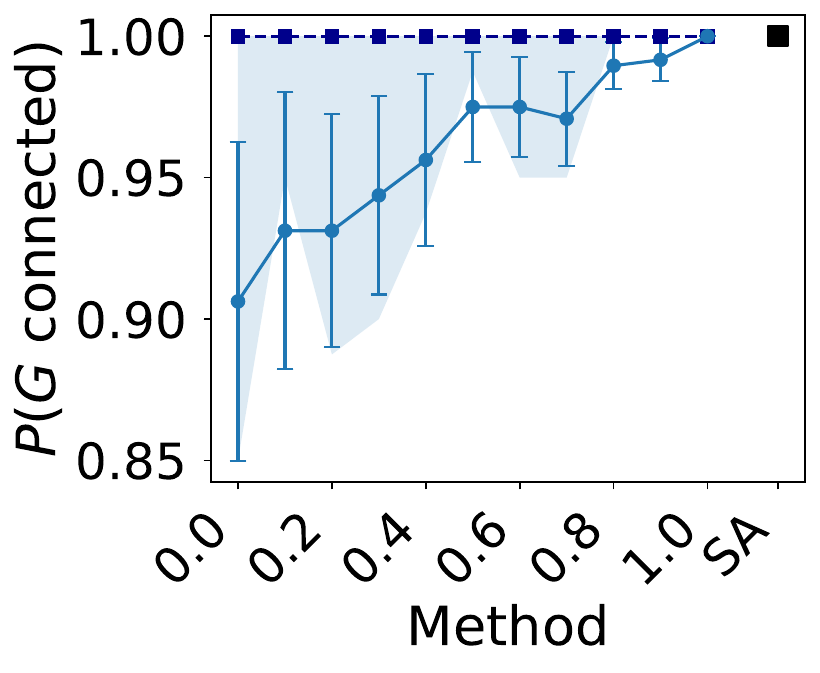}
\caption{Connectivity probability for small real-world networks (up to 500 nodes) as a function of $p$. We also include simulated annealing (SA). For each value of $p$, circles denote the mean probability of obtaining a connected graph across networks, with 95\% confidence intervals (CI) denoted by the corresponding error bars. Squares in the dashed line indicate the median, while the shaded region represents the interquartile range (IQR). 
}
\label{fig:sa_connectivity}
\end{figure}

\begin{figure}[t]
    \centering
\includegraphics[width=\columnwidth]{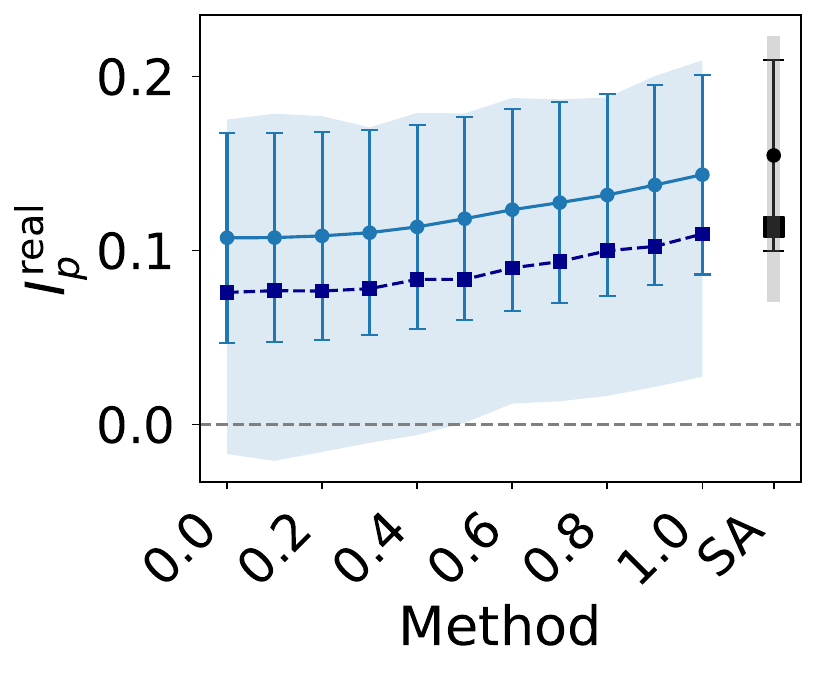}
\caption{Relative improvement in average shortest-path length with respect to the real networks for small networks, up to 500 nodes. Results are shown for different values of $p$ and for simulated annealing (SA). Circles denote the mean improvement across networks with 95\% confidence intervals (CI), the dashed line indicates the median, and the shaded region represents the interquartile range (IQR). Results are computed over networks for which all values of $p$ yield at least one connected realization.}
\label{fig:sa_rel_improv}
\end{figure}

\begin{figure}[t]
    \centering
\includegraphics[width=\columnwidth]{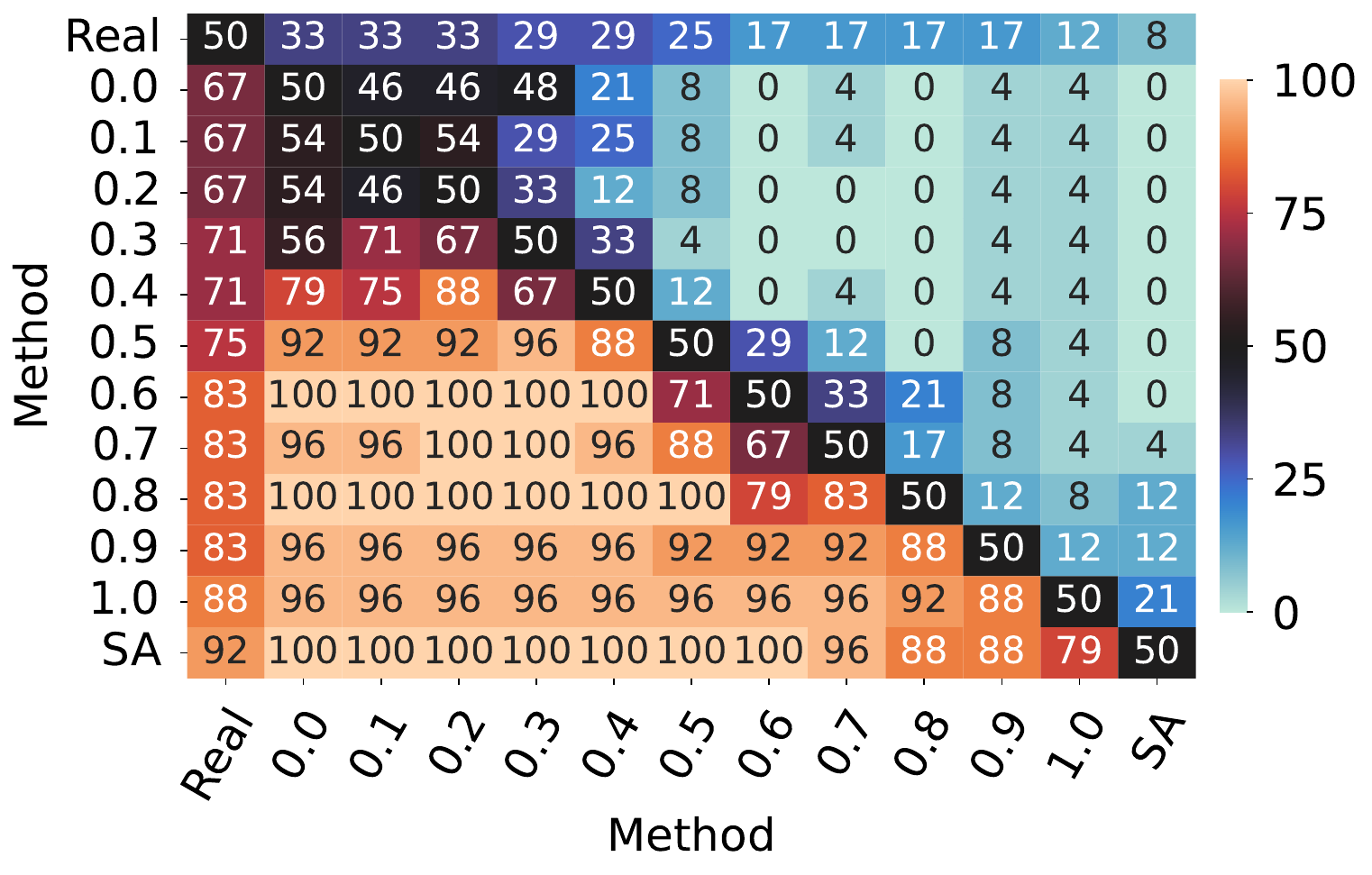}
\caption{Pairwise win percentage matrix on minimizing the average shortest-path length comparing DBCM for different values of $p$, simulated annealing (SA) and the real network, on networks with up to 500 nodes. Each entry $(i,j)$ reports the percentage of networks for which method $i$ achieves a smaller average shortest-path length than method $j$. Ties are counted as half a win for each method. Results are computed over connected network realizations.}
\label{fig:sa_pairwise}
\end{figure}

In Figure~\ref{fig:sa_connectivity}, we report the connectivity probability as a function of $p$, showing that for small networks the DBCM construction leads to connected networks in more than 90\% of cases, already for $p=0$. 
Figure~\ref{fig:sa_rel_improv} compares the relative improvement in average shortest-path length with respect to the real networks for small networks, up to 500 nodes, for the constructive method with different values of $p$ and for simulated annealing (SA).
Table~\ref{tab:sa_wins} summarizes the number of times each method achieves the minimum average shortest-path length and standard deviation. Finally, Figure~\ref{fig:sa_pairwise} presents the pairwise win percentage matrix, providing a detailed comparison between all methods, including simulated annealing and the real networks.

\begin{table}[ht]
\centering
\vspace{0.3em}
\begin{tabular}{
l@{\hspace{1.5em}}
S[table-format=2.0]
S[table-format=2.1]
S[table-format=2.0]
S[table-format=2.1]
}
\toprule
& \multicolumn{2}{c}{\textbf{$\mu$ Wins}} 
& \multicolumn{2}{c}{\textbf{Std Wins}} \\
\cmidrule(lr){2-3} \cmidrule(lr){4-5}
\textbf{Method} & {\textbf{Count}} & {\textbf{\%}} & {\textbf{Count}} & {\textbf{\%}} \\
\midrule
\textit{Real}    & 2  & 8.3  & 0  & 0.0 \\
$p=0.0$   & 0  & 0.0  & 0  & 0.0 \\
$p=0.1$ & 0  & 0.0  & 0  & 0.0 \\
$p=0.2$ & 0  & 0.0  & 0  & 0.0 \\
$p=0.3$ & 0  & 0.0  & 0  & 0.0 \\
$p=0.4$ & 0  & 0.0  & 0  & 0.0 \\
$p=0.5$ & 0  & 0.0  & 1  & 4.2 \\
$p=0.6$ & 0  & 0.0  & 1  & 4.2 \\
$p=0.7$ & 0  & 0.0  & 0  & 0.0 \\
$p=0.8$ & 0  & 0.0  & 0  & 0.0 \\
$p=0.9$ & 0  & 0.0  & 1  & 4.2 \\
$p=1.0$   & 5  & 20.8 
        & 9  & 37.5 \\
\textbf{SA} & \textbf{17} & \textbf{70.8}
            & \textbf{12} & \textbf{50.0} \\
\bottomrule
\end{tabular}
\caption{Number and percentage of small networks (up to 500 nodes) for which each method achieves the minimum average shortest-path length ($\mu$) or the minimum standard deviation (Std) of distances. $p$ is the parameter in the DBCM, \textit{Real} corresponds to the real network values and SA denotes simulated annealing. Results are computed over connected network realizations.}
\label{tab:sa_wins}
\end{table}

\section{Code Availability}
Code implementing the algorithms considered in this paper is available at \url{https://github.com/meritxell-vila/DBCM}.

\clearpage

\end{document}